\documentstyle[11pt,epsfig,here,a4]{article}
\begin{document}

{\bf DESY 99-073}
\vspace*{1.2cm}

\begin{center}

{\LARGE \bf The AMANDA Neutrino Telescope: }

{\LARGE \bf Principle of Operation and First Results}

\vspace{0.2cm}

%{\large \it The AMANDA Collaboration}
\end{center}
E.~Andres$^{10}$,
P.~Askebjer$^{4}$, 
S.W.~Barwick$^{6}$, 
R.~Bay$^{5}$, 
L.~Bergstr\"om$^{4}$, 
A.~Biron$^{2}$, 
J.~Booth$^{6}$, 
A.~Bouchta$^{2}$, 
S.~Carius$^{3}$, 
M.~Carlson$^{8}$, 
%C. Costa$^{8}$, 
D.~Cowen$^{7}$, 
E.~Dalberg$^{4}$, 
T.~DeYoung$^{8}$, 
P.~Ekstr\"om$^{4}$, 
B.~Erlandson$^{4}$,
A.~Goobar$^{4}$, 
L.~Gray$^{8}$, 
A.~Hallgren$^{11}$, 
F.~Halzen$^{8}$,  
R.~Hardtke$^{8}$,
S.~Hart$^{10}$, 
Y.~He$^{5}$,
H.~Heukenkamp$^2$,
G.~Hill$^{8}$, 
P.O.~Hulth$^{4}$, 
S.~Hundertmark$^{2}$, 
J.~Jacobsen$^{9}$,
A.~Jones$^{10}$, 
V.~Kandhadai$^{8}$, 
A.~Karle$^{8}$, 
B.~Koci$^{8}$, 
P.~Lindahl$^{3}$, 
I.~Liubarsky$^{8}$, 
M.~Leuthold$^{2}$, 
D.M.~Lowder$^{5}$, 
P.~Marciniewski$^{11}$, 
T.~Mikolajski$^2$,
T.~Miller$^{1}$, 
P.~Miocinovic$^{5}$, 
P.~Mock$^{6}$, 
R.~Morse$^{8}$, 
P.~Niessen$^{2}$, 
%D. Nygren$^{10}$, 
C.~P\'erez de los Heros$^{11}$, 
R.~Porrata$^{6}$,
D.~Potter $^{10}$, 
P.~B. Price$^{5}$, 
G.~Przybylski$^{9}$,
A.~Richards$^5$, 
S.~Richter$^{10}$,
P.~Romenesko$^{8}$,
H.~Rubinstein$^{4}$, 
E.~Schneider$^{6}$, 
T.~Schmidt$^{2}$, 
R.~Schwarz$^{10}$, 
M.~Solarz$^{5}$, 
G.M.~Spiczak$^1$,
C.~Spiering$^{2}$, 
O.~Streicher$^{2}$,
Q.~Sun$^{4}$,
L.~Thollander$^4$, 
T.~Thon$^{2}$, 
S.~Tilav$^{8}$, 
C.~Walck$^{4}$, 
C.~Wiebusch$^{2}$, 
R.~Wischnewski$^{2}$, 
K.~Woschnagg$^{5}$, 
G.~Yodh$^{6}$

{\it
1) Bartol Research Institute, University of Delaware, Newark, DE, USA \\
2) DESY-Zeuthen, Zeuthen, Germany \\
3) Kalmar University, Sweden \\
4) Stockholm University, Stockholm, Sweden \\
5) University of California Berkeley, Berkeley, CA, USA \\
6) University of California Irvine, Irvine, CA, USA \\
7) University of Pennsylvania, Philadelphia, PA, USA \\
8) University of Wisconsin, Madison, WI, USA \\
9) Lawrence Berkeley Laboratory, Berkeley, CA, USA \\
10) South Pole Station, Antarctica \\
11) University of Uppsala, Uppsala, Sweden \\}

\vspace{2mm}

{\bf Abstract:}
AMANDA is a high-energy neutrino telescope presently under
construction at the geographical South Pole. 
In the Antarctic summer 1995/96, an array of 80 optical 
modules (OMs) arranged on 4 strings (AMANDA-B4) was deployed at depths
between 1.5 and 2 km.
In this paper we describe the design and performance of the 
AMANDA-B4 prototype, based on data collected between February and 
November  1996. Monte Carlo simulations of the detector response
to down-going atmospheric muon tracks show that the global behavior
of the detector is understood. We describe the 
data analysis method and present  first
results on atmospheric muon reconstruction and separation
of neutrino candidates. 
The AMANDA array was upgraded  with 216 OMs on 6 new strings in 
1996/97 (AMANDA-B10), and 122 additional OMs on 3 strings in 1997/98.

\vspace{0.7cm}
\begin{center}
{\large \it Paper submitted to Astroparticle Physics}
\end{center}

\newpage

\section{Introduction}

Techniques are being developed by several groups to
use high energy neutrinos
as a probe for  the highest
energy phenomena observed in the Universe. Neutrinos yield information
complementary to that obtained from observations of high energy
photons and charged particles
since they interact only weakly
and  can  reach the observer unobscured by intervening matter
and undeflected by magnetic fields. 

The primary mission of large neutrino telescopes
is to probe the Universe in a new observational window and
to search for the sources of the highest
energy phenomena. Presently suggested candidates for these
sources are, for instance, Active Galactic Nuclei (AGN) 
and Gamma Ray Bursts (GRB).  A neutrino signal from a certain object
would constitute the clearest signature of the hadronic nature of 
that cosmic accelerator \cite{GHS}. Apart from that,
neutrino telescopes  search for neutrinos produced in annihilations
of Weakly Interacting Massive Particles (WIMPs) which may
have accumulated in the center of the Earth or in the Sun.
WIMPS might contribute to the cold dark matter content
of the Universe, their detection being of extreme importance
for cosmology \cite{WIMP}.
Neutrino telescopes can be also used to
monitor the Galaxy for supernova explosions \cite{SNHalzen}
and to search for exotic particles
like magnetic monopoles \cite{Mon}.
In coincidence with surface air shower arrays, deep neutrino
detectors can be used to study the chemical composition
of charged cosmic rays.
Finally, environmental investigations -- 
oceanology or limnology in water, glaciology in ice --
have proved to be exciting applications of these devices 
\cite{Baikal,Glac}.

Planned high-energy neutrino telescopes 
differ in many aspects from existing underground neutrino 
detectors.  Their
architecture is optimized to achieve a large detection
area rather than  a low energy threshold. 
They are  deployed in transparent "open" media like water
in oceans  or lakes, or deep polar ice.  This 
brings additional 
inherent technological challenges compared with the assembly of a detector  
in an accelerator tunnel or underground cavities.
Neutrinos are inferred from the arrival times of Cherenkov
light emitted by charged secondaries produced in  neutrino
interactions.  
The light is mapped by photomultiplier tubes 
(PMTs) spanning a  coarse three-dimensional  grid. 

The traditional approach to muon neutrino detection is the observation
of upward moving muons produced in  charged current interactions in the
rock, water or ice below the detector.
The Earth is used as a filter with respect to atmospheric muons. 
Still, suppression of
downward-going muons is of top importance, since their flux exceeds 
that of upward-going muons from atmospheric neutrinos 
by several  orders of magnitude. 

An array of PMTs can also
be used to reconstruct the energy and location of isolated cascades due
to neutrino interactions. Burst-like events, like the onset
of a supernova, might be detected by measuring the increased
count rates of all individual PMTs.

Technologies for under{\it water} telescopes have been pioneered
by the since  decommissioned DUMAND 
project near Hawaii \cite{DUMANDWWW, DUMAND} and
by the Baikal collaboration \cite{Baikal,BAIKALWWW}.
In contrast to these approaches, the AMANDA detector \cite{Am0}
uses deep polar ice as target and radiator.
Two projects in the Mediterranean,
NESTOR \cite{NESTORWWW} and ANTARES \cite{ANTARESWWW}, have joined the
worldwide effort towards large-scale underwater telescopes. 
BAIKAL and AMANDA are presently taking data with first stage
detectors.

The present paper describes results obtained with the first four
(out of the current thirteen) strings of the AMANDA detector. 
The paper is organized as follows: In section~\ref{concept} we give a general 
overview of the AMANDA  
concept. Section~\ref{amandaa} summarizes the results obtained with a shallow
survey detector called AMANDA-A. Section~\ref{deployment} describes the design
of the first four strings of the deeper array  AMANDA-B4.
Calibration of time response and of geometry are explained in section 5.
In section~\ref{simureco} we describe the simulation and
reconstruction methods with respect to atmospheric muons and compare
experimental data to Monte Carlo calculations.
Section~\ref{spase}  demonstrates the performance of AMANDA-B4 operated in
coincidence with SPASE, a surface air shower array.
In section~\ref{depth}, the angular spectrum 
of atmospheric muons  is derived and transformed into
a dependence of the vertical intensity on depth.
Section~\ref{upward} describes the separation of first 
upward going muon candidates.
Finally, a summary of 
the status of AMANDA and results is presented in 
section~\ref{conclusion}.

%\newpage

\section{The AMANDA Concept \label{concept}}

\begin{figure}[htbp]
\centering
\hspace{0.5cm}
\mbox{\epsfig{file=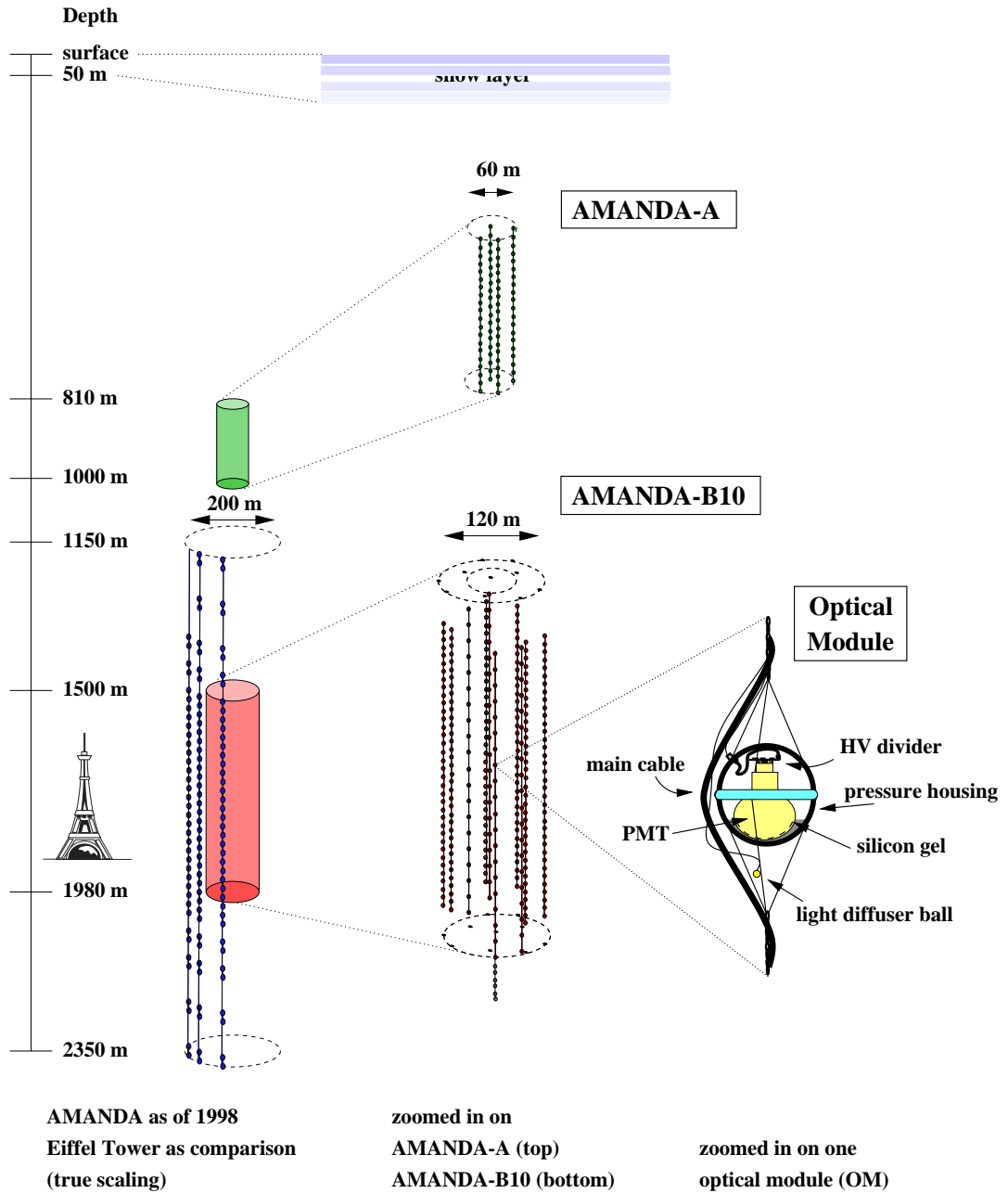,height=17.0cm}}
              \caption {\small \label{fullamanda}
      Scheme of the 1998 AMANDA installations. The left picture  is
drawn with true scaling. A zoomed view on AMANDA-A
(top) and AMANDA-B10 (bottom) is shown at the center. The right zoom 
depicts the optical module.
      }
\end{figure}

AMANDA (Antarctic Muon And Neutrino Detector Array) uses the natural
Antarctic ice as both target and Cherenkov medium. 
The detector consists of  
strings of optical modules (OMs) frozen in the 
3 km thick ice sheet at the South Pole. An OM consists of a
photomultiplier in a 
glass vessel. The strings are 
deployed into holes drilled with pressurized hot water. The 
water column in the hole then refreezes within 35-40 hours, fixing 
the string in its final position. In our basic design, each OM has 
its own cable supplying the high voltage (HV)
as well as transmitting the anode signal. 
The components under the ice are kept as simple as possible, all the 
data acquisition electronics being housed in a building at the surface. 
The simplicity of the components under ice and the non-hierarchical
structure make the detector highly reliable.

Fig.~\ref{fullamanda} shows the current configuration of the AMANDA detector.
The shallow array, AMANDA-A, was deployed at a depth 
of 800 to 1000\,m
in 1993/94 in an exploratory phase of the project.
Studies of the optical properties of the ice  
carried out with AMANDA-A showed that a high 
concentration of residual air bubbles remaining at these depths  
leads to strong scattering of light, making 
accurate track reconstruction impossible \cite{Aske}. 
Therefore, in the polar season
1995/96 a deeper
array consisting of 80 OMs arranged on four strings
(AMANDA-B4) 
was deployed at depths ranging from 1545 to 1978 meters, where the 
concentration of bubbles was predicted to be negligible according to  
extrapolation of AMANDA-A results. 
The detector was upgraded in 1996/97 with 216 additional OMs on
6 strings. This detector of 4+6 strings was named AMANDA-B10 and is sketched
at the right side of fig.~\ref{fullamanda}. AMANDA-B10 was upgraded in the 
season 1997/98 by 3 strings instrumented between 1150\,m and 2350\,m 
which fulfill several
tasks. Firstly, they  explore the very deep and very shallow ice
with respect to a future cube kilometer array. Secondly, they
form one corner of AMANDA-II which is the next stage of AMANDA
with altogether about 700 OMs. Thirdly, they have been used
to test data transmission via optical fibers.

There are several advantages that make the 
South Pole a unique site for a neutrino telescope:

\begin{itemize}

\item The geographic location is unique:
        A detector located at the  South Pole
         observes the northern hemisphere, and
        complements any other of the planned or existing detectors.

\item Ice is a sterile medium.
      The noise is given only by the PMT dark
      noise and by $K^{40}$ decays in the glass housings, 
      which are 0.5-1.5 kHz for the PMTs and spheres we used. 
      Ocean and lake experiments
      have to cope with 100 kHz noise rates due to bioluminescence
      or $K^{40}$ decays (25-30 kHz if normalized to the photocathode 
      area of the 8$^{\prime \prime}$ PMT used in AMANDA). 
      This fact not only facilitates counting rate experiments
      like the search for 
      low energy neutrinos from supernovae or GRBs,
      but also leads to fewer accidental hits 
      in muon  events -- an essential 
      advantage for trigger formation and track reconstruction.

\item AMANDA can be operated in coincidence with air shower arrays
      located
      at the surface. Apart from complementing the information from
      the surface arrays by measurements of muons penetrating to 
      AMANDA depths, the air shower information can be used to
      calibrate AMANDA.

\item The South Pole station has an excellent infrastructure. Issues
    of vital importance to run big experiments like transportation,
   power supply, satellite communication and technical support are solved
   and tested during many years of operation. 
   Part of an existing building can be used to house the surface electronics.

\item The drilling and deployment procedures are
tested and well under control. AMANDA benefits from the 
drilling expertise of the Polar Ice Coring Office (PICO).  
Currently about 
five days are needed to drill a hole and to deploy a string with PMTs
to a depth of 2000\,m. Future upgrades of the drilling
equipment are expected to result in a further speed-up.

\end{itemize}

The optical properties of the ice turned out to be very
different from what had been expected before the AMANDA-A phase. 
Whereas absorption is
much weaker than in oceans, scattering effects turned out to
be much stronger. Even at depths below 1400 meters, 
where residual bubbles  have collapsed almost completely
into air hydrates,
scattering is nearly an order of magnitude stronger than in water
(see below). 
Since scattering of light smears out the arrival
times of Cherenkov flashes, a main question was whether under
these conditions track reconstruction 
was possible. As shown below, the answer is yes.
%The most important conclusion of this paper is that ice is an 
%adequate medium  
%to do muon track reconstruction in the presence of scattering. 

%\newpage

\section{AMANDA-A: A First Survey \label{amandaa}}

Preliminary explorations of the site and the drilling technology
were performed in the Antarctic Summer 1991/92 \cite{Am0}.
During the 1993/94 campaign, four strings each carrying 20 OMs 
("AMANDA-A") were  deployed between 800 and 1000\,m
depth. None of the 73
OMs (equipped with 8$^{\prime \prime}$ EMI PMTs)
surviving the refreezing process failed during the following two
years, giving a mean time between failures (MTBF) $>$ 40 years for
individual OMs in AMANDA-A. 
The OMs are connected to the surface electronics by 
coaxial cables. 
Along with the coaxial cables, 
optical fibers  carry light  from a Nd:YAG laser at the surface to 
nylon light diffusers placed about 30\,cm below each PMT
(see fig.~\ref{fullamanda}).
Time calibration is performed by  
sending nanosecond laser pulses to individual diffusers and measuring the
photon arrival time distributions at the closest  PMT.
From the distribution of the arrival times at {\it distant} PMTs, 
the optical
properties of the medium were derived \cite{Aske,Glac}. 
The measured timing distributions indicated that photons do not propagate
along straight paths
but are scattered and considerably delayed due to
residual bubbles in the ice. The distributions
could be fitted well with an analytic function describing the
three-dimensional random walk (scattering) including absorption.
These results showed that polar ice at these depths
has a very large  absorption
length, exceeding 200\,m at a wavelength of 410\,nm.
Scattering is described by the effective
scattering length $L_{eff} = L_{sc}/ (1 - \langle \cos \theta
\rangle)$,
where $L_{sc}$ is the geometrical scattering length and
$\langle \cos \theta \rangle$ the average cosine of the scattering
angle \cite{Aske}.  $L_{eff}$
increases with depth, from 40\,cm at 830\,m
depth to 80\,cm at 970\,m.
In accordance with measurements
at the Vostok Station (East Antarctica \cite{Vostok}) 
and Byrd Station (West Antarctica)
these results suggested that at depths greater than 1300-1400\,m
the phase transformation from bubbles into
air-hydrate crystals would be complete and bubbles would 
disappear.

Although not suitable for track reconstruction, AMANDA-A 
can be used as a calorimeter for energy measurements of 
neutrino-induced cascade-like events \cite{Rodin}. It is also used as 
a supernova monitor \cite{Ralf}. Events that simultaneously trigger
AMANDA-A and the deeper AMANDA-B have been used for methodical
studies like the investigation of the optical
properties of the ice or the assessment of events with a lever
arm of one kilometer.

%\newpage

\section{Deployment and Design of AMANDA-B4 \label{deployment}}

\subsection{{\bf Drilling and} Deployment Procedure}

Drilling is performed by melting the ice with pressurized 
water at 75$^o$C. The drilling equipment 
operates at a  power of 1.9 MW and the typical drill speed is about 1 cm/s. 
It takes about 3.5 days 
to drill a 50-60\,cm diameter hole to 2000\,m depth.
% since the drilling speed 
%is reduced with depth.

In the season 1995/96, we drilled four holes,
the deepest of them reaching 2180\,m.
It took typically 8 hours to remove the drill and the
water recycling pump from the completed hole.
The deployment of one string
with 20 OMs and several calibration devices
took about 18 hours (with a limit of 35 hours set by the
refreezing of the water in the hole).

Several diagnostic devices allow  
monitoring of the mechanical and thermal parameters during the entire 
refreezing process and
afterwards. 
%Four inclinometers measured shear vs. time, thermistors
%measured temperature vs. depth and 
%pressure gauges followed the refreezing. 
It was shown that the temperature increases with
depth in good agreement with the prediction of a standard
heat flow calculation for South Pole ice.
At the greatest depth, the temperature of the ice is 
$\approx$ -31$^o$C, about 20$^o$ warmer than at the surface.
During the refreezing, the pressure reached a maximum of 460 atm,
more than twice the hydrostatic pressure which is asymptotically
established. 

\subsection{Detector Design}

The four strings of AMANDA-B4 were deployed 
at depths between 1545 and 1978\,m.
An OM consists of a 30 cm diameter glass sphere 
equipped with a 8$^{\prime \prime}$ Hamamatsu R5912-2 photomultiplier, 
a 14-dynode version of the standard 12-dynode R5912 tube. 
The PMTs are operated at a gain of 10$^9$ in order to drive the 
pulses through 2\,km of coaxial cable without in-situ amplification. 
The amplitude of a one-photoelectron pulse is about 1 V.
The coaxial cable is also used for the HV supply, with the advantage
that only one cable and one electrical penetrator into the
sphere are required for each OM.
The measured noise rate of the AMANDA-B4 PMTs is  typically 400 Hz
(threshold 0.4 photoelectrons). 

The photocathode is in 
optical contact with the glass sphere by the use of silicon gel. 
The transmission of the glass of the pressure sphere
is about 90\% in the spectral range between 400 and 600\,nm;
the 50\% cutoff on the UV side is at about 365 nm.
The glass spheres are designed to withstand pressures of 
about 660 atm.

\begin{center}
\begin{figure}[htbp]
\centering
\hspace{0.5cm}
\mbox{\epsfig{file=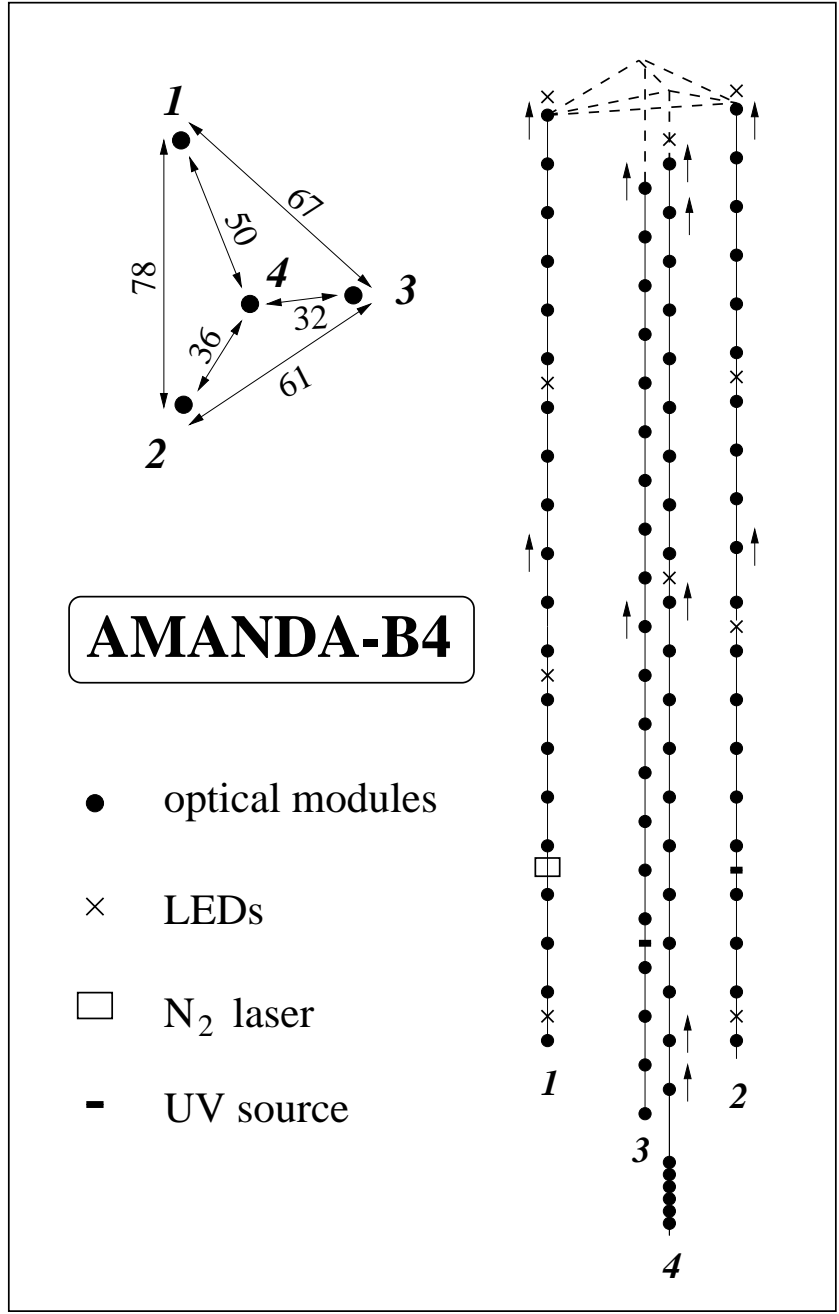,height=17.5cm}}
              \caption {\small 
AMANDA-B4: Top view, with distances 
between strings given in meters, and side view
showing optical modules and calibration light sources. Upward
looking PMTs are marked by arrows.}
\label{b4_design}
\end{figure}
\end{center}

Each string carries 
20 OMs with a vertical spacing of 20\,m. The fourth string 
carries six additional OMs connected by a twisted pair cable.
These six OMs will not be used in the analyses presented in this
paper.

Fig.~\ref{b4_design} shows a schematic view of  AMANDA-B4.
All PMTs look down with the exception of \# 1,10  in strings 1 to
3 and \#1,2,10,19,20  in string 4 (with the numbers running
from top to bottom of a string). Strings 1-3 
form a triangle with side lengths 77-67-61\,m; string 4 is close to
the center. 
The OMs are arranged at  depths 1545--1925\,m (string 1),
1546--1926\,m (string 2),
1598--1978\,m (string 3) and
1576--1956\,m (string 4). The  additional six OMs equipped 
with twisted pair cables are at string 4 between 2009 and 2035\,m.
Seven of the 80 PMTs which define AMANDA-B4 
were lost due to overpressure and 
shearing forces to the electrical connectors during
the refreezing period.
These losses can be reduced by computer controlled
drilling avoiding strong irregularities in the hole
diameter, and by
improved connectors. Another 3 PMTs failed  in the course of
the first 3 years of operation, giving a MTBF of 73 years.
%This rate might hopefully be reduced by operating the PMT
%at lower voltage which is planned in connection with novel
%read-out schemes. 

\subsection{Electronics and DAQ}

\begin{center}
\begin{figure}[htbp]
\centering
\hspace{0.5cm}
\mbox{\epsfig{file=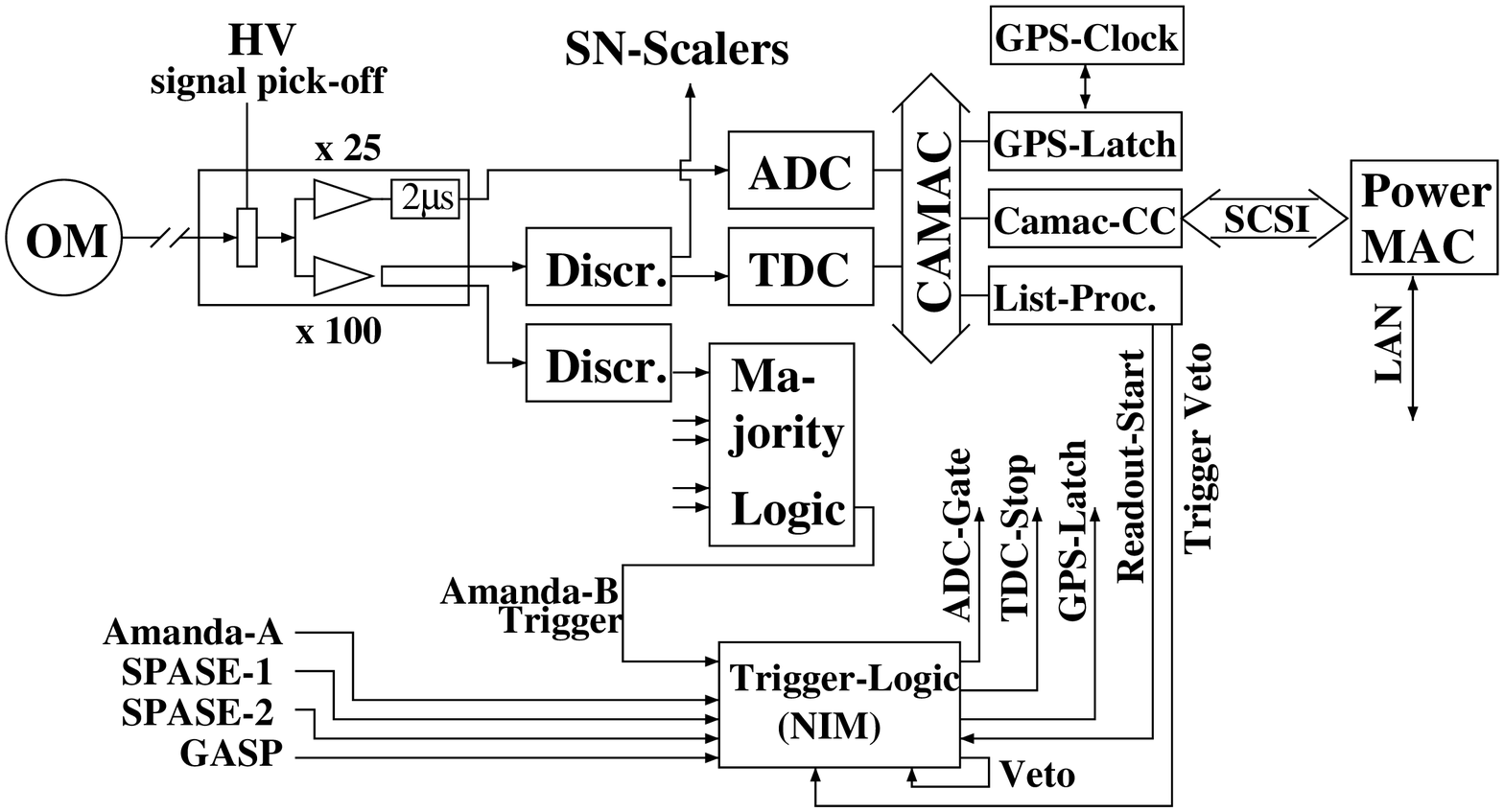,height=8.5cm}}
              \caption {\small 
DAQ system used for AMANDA-B4 during 1996
}
\label{DAQ}
\end{figure}
\end{center}

Each PMT can give a series of pulses which can be resolved if
separated from each other by more than a few hundred nanoseconds.
The data recorded consist of the leading and trailing edges of
the pulses. The time-over-threshold gives a measure of the
amplitude of individual pulses. Another measure of the amplitude is
obtained by a voltage sensitive ADC which records the peak value
out of the subsequent hits of an event in a PMT. Actually, 
the information consists of leading and trailing edges
of the last 8 resolved pulses, and of the largest amplitude of
those of them which lie in a 4\,$\mu$sec window centered at the
array trigger time. Also recorded is the GPS time at which the event
occurred. 
A scheme of the AMANDA electronics layout is shown in fig.~\ref{DAQ}.

The signal from each
cable is fed to a module  consisting of a DC blocking high-pass 
filter which picks up the pulse, a fan-out sending it 
to 2 amplifiers with 100$\times$  and 25$\times$ gain,
and a 2 $\mu$sec delay for the low-gain signal.

The delayed signal is sent to a Phillips 7164 peak sensing ADC.
The other pulse is split and sent to LeCroy 4413 discriminators with
thresholds set at 100 mV corresponding to about 0.3-0.4 photoelectrons
at the given high voltage.
One of the resulting ECL pulses is fed into a LeCroy 3377 TDC while 
the other is sent to the majority trigger. The TDC records
the last 16 time edges occurring within a 32 $\mu$sec time window.

The majority logic requests $\ge$ 8 hit PMTs within a sliding window
of 2 $\mu$sec. The trigger produced by this majority scheme
is sent to the NIM trigger logic. The latter
accepts also triggers from 
AMANDA-A or the air shower experiments SPASE-1, SPASE-2 and GASP. 
Thus AMANDA also records data when these detectors trigger even 
if a proper AMANDA trigger is not fulfilled.
The total trigger
rate during 1996 was about 26 Hz on average. The coincidences
from the other detectors contributed about 8 Hz to the total rate. 

The differences 
in cable length are not compensated before triggering. Therefore 
the true trigger window would be about 300 nsec for a vertically
downgoing relativistic particle and $\approx 4 \mu$sec for
an upgoing one. As a result downgoing particles are suppressed 
compared to upgoing.

Upon triggering, an ADC gate of 4\,$\mu$sec width 
is formed, a stop signal is sent to
the TDCs and a readout signal is sent to a Hytec LP1341 list
processor.
Then a veto lasting several microseconds inhibits
further trigger signals.

A separate system ("SN scalers" in fig.~\ref{DAQ}) monitors the
counting rates of individual PMTs and searches for rate excesses
lasting several seconds. Such an increase would be expected for
multiple low-energy neutrino interactions close to each PMT due to 
a supernova burst \cite{SNHalzen,Ralf}.

The AMANDA-B4 DAQ was running on a MacIntosh Power PC communicating 
through a SCSI bus with the CAMAC crate controller. 
From the distribution of the time differences between
subsequent events, the dead time of the DAQ is
estimated to be about 12\,\%. The MacIntosh has
been replaced by a Pentium-II PC running under LINUX in 1998, and 
part of the CAMAC electronics by VME modules.

\begin{center}
\begin{figure}[htbp]
\centering
\hspace{0.5cm}
\mbox{\epsfig{file=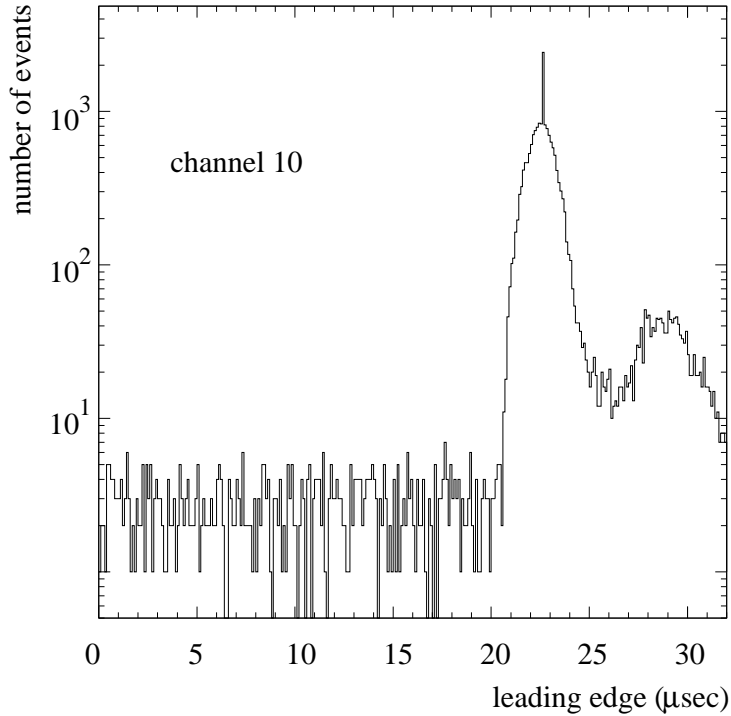,height=9.5cm}}
              \caption {\small 
Leading edge times of PMT \# 10 of AMANDA-B4 for data
taken with an 8-fold majority trigger.
}
\label{LE}
\end{figure}
\end{center}

Fig.~\ref{LE} shows the distribution of the leading-edge times
of one PMT for data taken with the 8-fold majority trigger.
The sharp peak at 23 $\mu$sec is given by the
time when this PMT was the triggering one (i.e. the eighth) within a 
2$\mu$sec window.
The flat part is due to noise hits and the bulge after the
main distribution to afterpulses (about 6\%.)

\subsection{Calibration Light Sources and Ice Properties}

An essential ingredient to the operation of a detector like 
AMANDA is the  knowledge of the optical properties of the 
ice, as well as a precise time calibration of the detector. 
Various light calibration sources have been deployed at 
different depths in order to tackle these questions:

\vspace{12mm}
 
\begin{itemize}
 
\item {\bf The YAG laser calibration system}. It uses optical fibers 
with diffusers located at each PMT. This system is
similar to that used for AMANDA-A.
The range of transmittable wavelengths is $\ge$ 450\,nm, the
time resolution is
about 15\,nsec at 530\,nm, the maximum intensity 
emitted by the diffusers is $10^{8}$ 
photons/pulse. Apart from ice investigations, the
laser system is used for time calibration of the PMT closest
to the diffuser  and for position calibration (see section~\ref{calib_time_geo}).
 
%\vspace{-2mm}
 
\item {\bf A nitrogen laser} at 1850\,m depth, wavelength 337\,nm,
pulse duration 1\,nsec, with a maximum intensity of
$10^{10}$  photons/pulse.
 
%\vspace{-2mm}
 
\item {\bf Three DC halogen lamps} (one broadband and two with
filters for 350 and 380\,nm), maximum intensity  $10^{14}$ 
(UV-filtered) 
and $10^{18}$ (broadband) photons/second.
 
%\vspace{-2mm}
 
\item {\bf LED beacons}, operated in  pulsed mode (500 Hz, 
pulse duration  7~nsec, $10^6$ 
photons/pulse) and DC mode ($10^{14}$ to $10^{15}$ photons/sec), wavelength
450\,nm. A filter restricts the output of a few beacons 
to 390\,nm, with reduced intensity.
 
%\vspace{-2mm}

\end{itemize}

Time-of-flight  measurements have been made for 
a large variety of combinations of optical fiber
emitters and PMTs for the YAG laser system, 
and at different wavelengths and intensities.
The nitrogen laser provided data at 337 nm.
The result is a considerable data base of hundreds of time distributions.
The width of the distributions is sensitive predominantly
to scattering and the tail to absorption (see 
\cite{desyproposal} for details).
The DC sources provide data for attenuation, i.e.
the combined effect of absorption and scattering.

\begin{center}
\begin{figure}[htbp]
\centering
\hspace{0.5cm}
\mbox{\epsfig{file=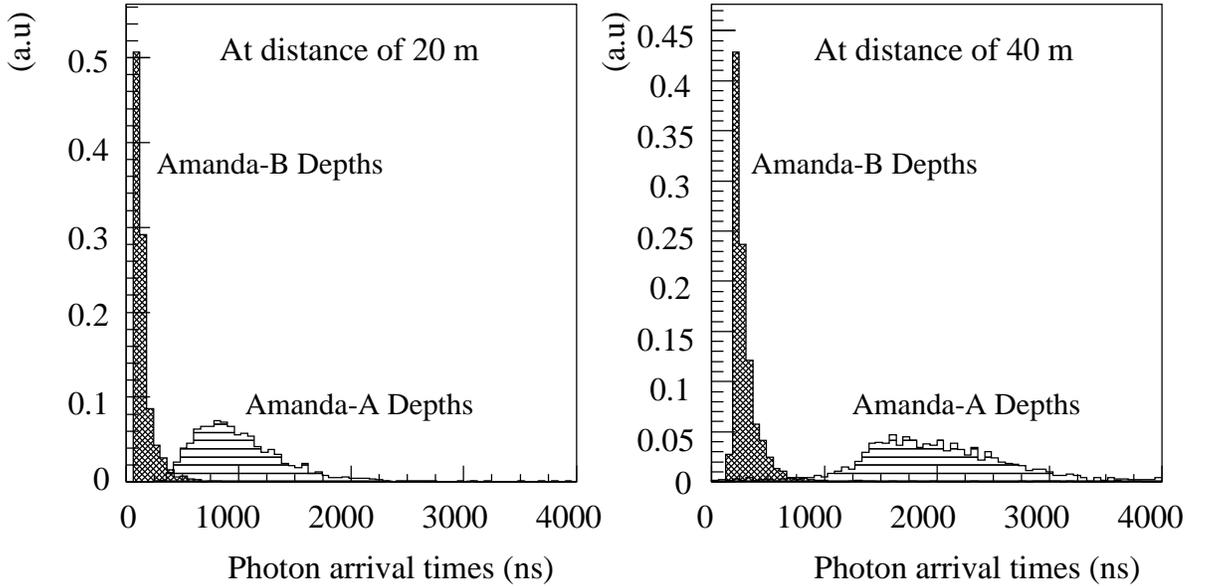,height=7.7cm}}
              \caption {\small
Arrival time distributions for 510 nm photons for two source-detector
distances. Black histograms: AMANDA-B. Hatched histograms:AMANDA-A. 
The histograms are normalized to the same area.        
}
\label{A-B-comparison}
\end{figure}
\end{center}

The YAG laser results  indicate  a 
dramatic improvement compared to AMANDA-A  results. 
Fig.~\ref{A-B-comparison} 
shows the distributions of arrival time for source-detector 
distances of 20 and 40 m, respectively,
for AMANDA-A as well as  AMANDA-B depths. The much smaller widths
for AMANDA-B support the expectation that bubbles as the dominant 
source of
scattering have mostly disappeared at
depths between 1550 and 1900\,m \cite{Vostok}.

\vspace{8mm}

\vspace{5mm}

\begin{center}
\begin{figure}[htbp]
\centering
\hspace{0.5cm}
\mbox{\epsfig{file=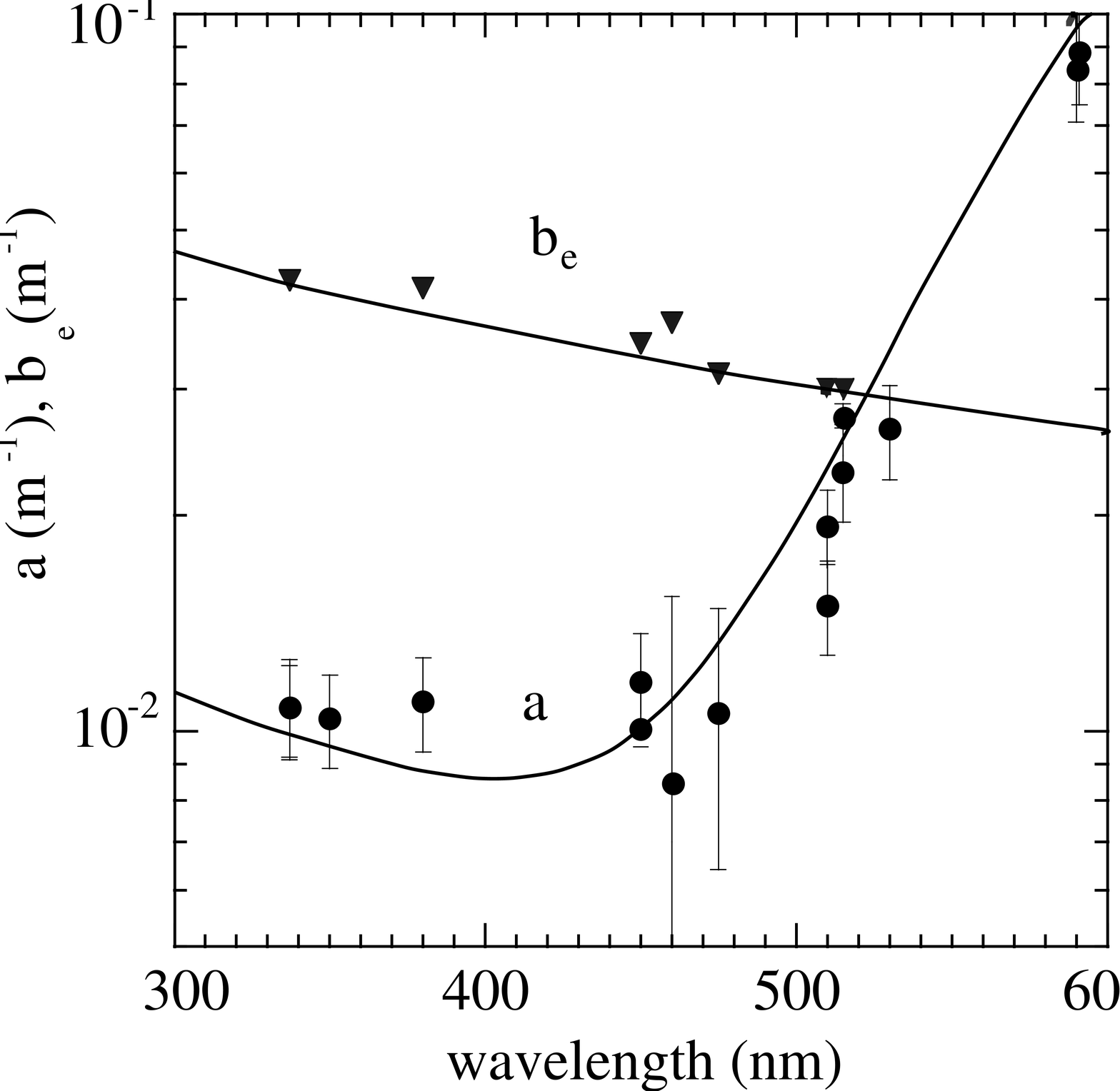,height=9.7cm}}
              \caption {\small
Absorption ($a$) and scattering ($b_e$) coefficients  
at an average depth of 1.7\,km,
compared to theory of He and Price \cite{He}.
}
\label{He}
\end{figure}
\end{center}

Details of the analysis of the optical properties of the 
ice at AMANDA-B4 depths 
have been  published elsewhere  \cite{Kurt}.  
Final results will be published in a separate paper.
Figure \ref{He} shows preliminary data on the wavelength 
dependence of the coefficients for scattering, $b_e$, 
and absorption, $a$. The absorption length $\lambda_a = 1/a$ 
is between 90 and 100~m
for wavelengths below 460~nm, i.e. ice does not degrade in
transparency towards smaller wavelengths down to 337 nm. 
The effective scattering length 
$\lambda_{eff} = 1/b_{e}$  
varies
between 24 and 30 m in the relevant wavelength range.
$\lambda_{eff} = \lambda_{scatt}/(1 -  \langle \cos{\theta} \rangle)$,
with $\lambda_{scatt}$ being the geometric scattering length. 
$\langle \cos \theta \rangle $ is the average cosine of the
scattering angle and is supposed to be about 0.8 in deep ice.
The attenutation length $\lambda_{att}$ 
which characterizes 
the decrease of the photon flux as a function of the distance
is about 27\,m.
These values are averages over the full
depth interval covered by AMANDA-B4. The variation 
of attenuation over this
depth range is within $\pm 30 \%$.

\vspace{-5mm}

\section{Calibration of Time Response and Geometry \label{calib_time_geo}}

\subsection{Time Calibration}

The measured arrival times from each PMT have to be corrected for
the time offset $t_0$, that is, 
the time it takes  a
signal to propagate through the PMT and the coaxial cable and get
digitized by the DAQ. 
The time offset is determined 
by sending light pulses from the  YAG laser 
to the diffuser nylon balls
located below each OM.  Two  fibers are available
for each PMT, one single and one multi-modal.
The time it takes for light to travel though the fiber is measured
using an OTDR  (Optical Time Domain Reflectometer)  and
subtracted from the time distributions recorded. 
 
For each PMT, the time difference between the laser
pulse at the surface and the PMT response arriving back is measured.
Upon arrival at the surface, the pulses have traveled through nearly
2000 meters of cable and are dispersed, with
typical time-over-thresholds of 550 nsec and rise times of 180 nsec.
The threshold used for TDC measurements is set to a constant value
with the consequence that small pulses will reach that value later
than larger ones.  This causes an amplitude-dependent offset 
or  "time walk",  which can be corrected for by

\begin{equation}
\label{eq:adc_correction}
t_{true} = t_{LE} - t_0 - \alpha / \sqrt{ADC}. 
\end{equation}

\begin{figure}[htbp]
\begin{center}
%\mbox{\epsfig{figure=adc-correction.eps,width=0.\textheight}}
\mbox{\epsfig{figure=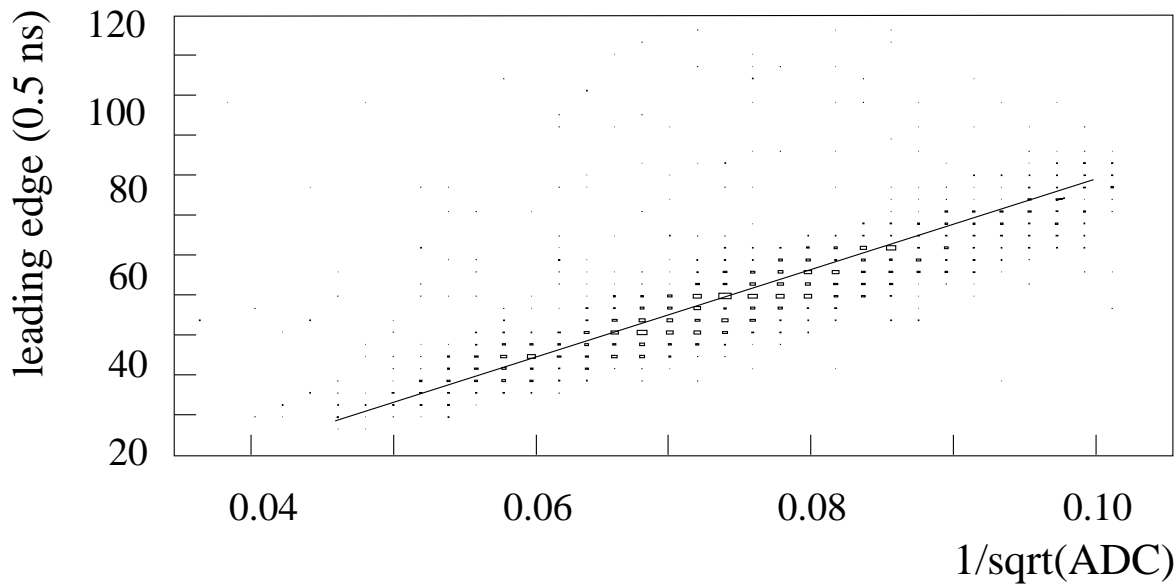,width=15.5cm}}
\caption{\small {Example of a fitted leading edge (with
100$<$ADC$<$1200) for module 19 on string 3. The ADC value
measures the peak value of the amplitude.}}
\label{adc-correction}
\end{center}
\end{figure}

Here, $t_{LE}$ is the measured leading edge
time and $t_{true}$ the true time at
which the light pulse reaches the photocathode.
The estimates of the time offset  $t_0$ and the time-walk
term $\alpha$ are extracted from scatterplots like
the one shown in fig.~\ref{adc-correction}.

\begin{figure}[htbp]
\begin{center}
\mbox{\epsfig{figure=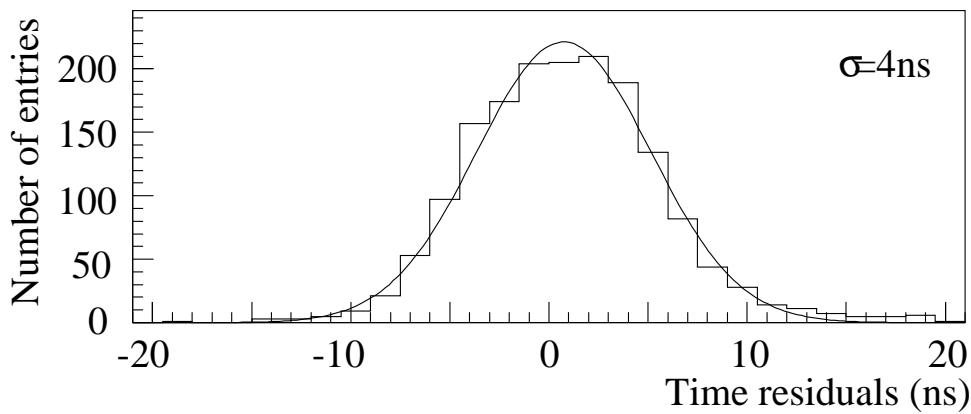,width=13cm}}
\caption{\small {Residuals after subtracting the time
correction obtained with the fitted parameters 
$t_0$ and $\alpha$ for module 19 on string
3. The standard deviation of the Gaussian fit is 4 nsec.}}
\label{time-resolution}
\end{center}
\end{figure}

The time resolution achieved in this way can be 
estimated by the standard deviation of a Gaussian fit to the
distribution of time residuals after correction, yielding 4--7\,nsec (see
Fig.~\ref{time-resolution} for an OM with 4\,nsec resolution).
Part of the variation is due to quality variations of the 
1996 optical fibers. Laboratory measurements yield a Gaussian 
width of 3.5 nsec after  2\,km cable.

%A calibration of the whole array was performed in the 1995/96 season
%and repeated one year later, with a standard deviation of about
%6 nsec in $t_0$ between the two sets. 
%Most of this difference can be assigned to larger
%systematic errors in the OTDR measurements in the 1995/96 
%calibration which had been performed with light pulses at
%850\,nm wavelength. Compared to the 532\,nm YAG pulses this
%resulted in a slightly different index of refraction of the fiber,
%and a correspondingly different transmission speed which had to be
%corrected for. 
%This error is reduced in
%The new OTDR operates close to 530\,nm wavelength.

%We therefore conclude, that the typical
%accuracy of the time calibration after proper OTDR corrections
%is 5-8\,nsec.

\subsection{Position Calibration}

Information about the  exact geometry of the array can be
obtained by different methods. Firstly, the measured propagation times
of photons between different light emitters and receivers 
can be used to determine their relative positions. Secondly,
absolute positions can be obtained from drill recordings
and pressure sensors.

\medskip

{\large \it Laser Calibration}

\begin{figure}[htbp]
\begin{center}
%\mbox{\epsfig{figure=bias_measurement.eps,width=0.65\textheight}}
\mbox{\epsfig{figure=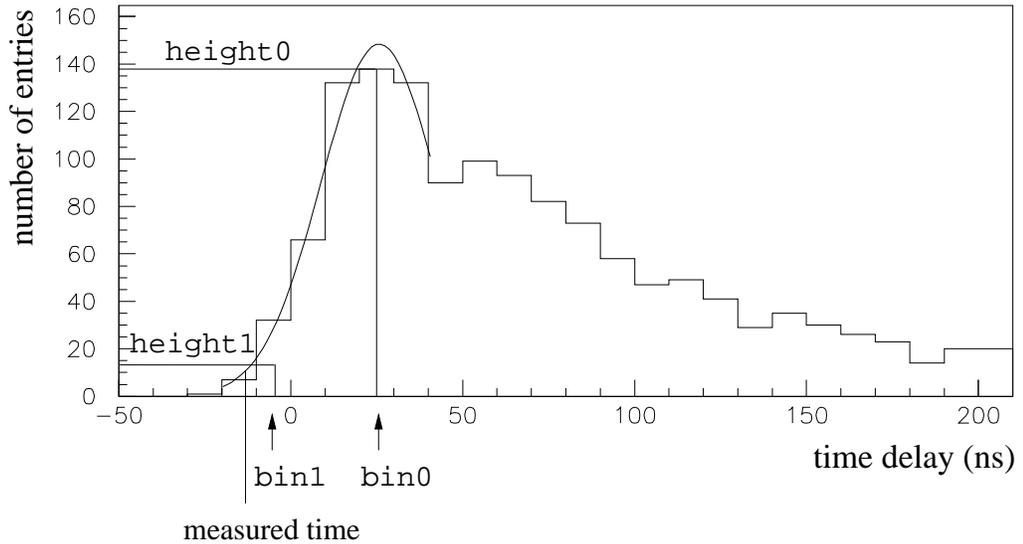,width=13.5cm}}
\caption{\small {Simulated time-shift distribution for 1500 
one-photoelectron events, for a distance of 60\,m
between emitter and receiver.
A Gaussian smearing of 10\,nsec
was applied to individual entries. Clear ice would yield a 
10\,nsec wide peak at 0\,nsec.}}
\label{bias}
\end{center}
\end{figure}

The YAG laser, the nitrogen laser and the pulsed LEDs can be used 
to infer the OM positions from the time-of-flight of photons
between these light sources and  the OMs. The zero time is
determined from the response of the OM closest to the light source
which is triggered by unscattered photons. This PMT is lowered in 
voltage in order not to be driven in saturation, and a
time correction accounting for the longer PMT transit time
is added.
In contrast to the close OM, 
the distant OMs  see mostly scattered photons.
However, for a few of the events out of a series
of about 1500 laser pulses, the
leading edge should be produced by
photons which are only slightly scattered.
Therefore the distance between  emitter and OM can be estimated
from the earliest events in the time-difference distribution (see~fig.~\ref{bias}). 

In order to reduce the sensitivity to fluctuations in the number of
early hits and binning effects, the whole left flank of the distribution is
fitted with a Gaussian between the maximum of the distribution
({\tt height0} in fig.~\ref{bias}) and the first bin with 
a height larger than {\tt height1} =1/10 {\tt height0}. 
The corrected "first" time is given by that bin ({\tt bin1})
for which the
fitted Gaussian yields a height exceeding {\tt height1}.
This time  has to be corrected further for the shifts due to
scattering which are expected even for the first bin of the
distribution. The corrections were obtained from 
Monte Carlo (MC) calculations
and are almost insensitive to variations in absorption and scattering 
length of a few meters.

\begin{figure}[htbp]
\begin{center}
\mbox{\epsfig{figure= 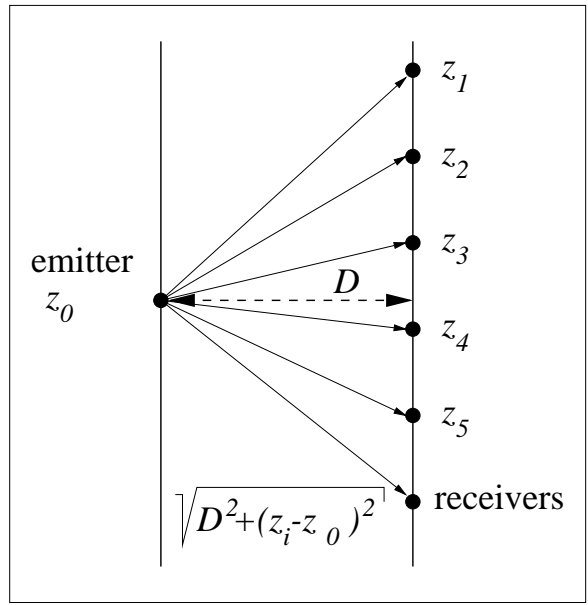,height=8cm}}
\caption{\small {Principle of position measurement}}
\label{pos_princip}
\end{center}
\end{figure}

\begin{figure}[htbp]
\begin{center}
\mbox{\epsfig{figure=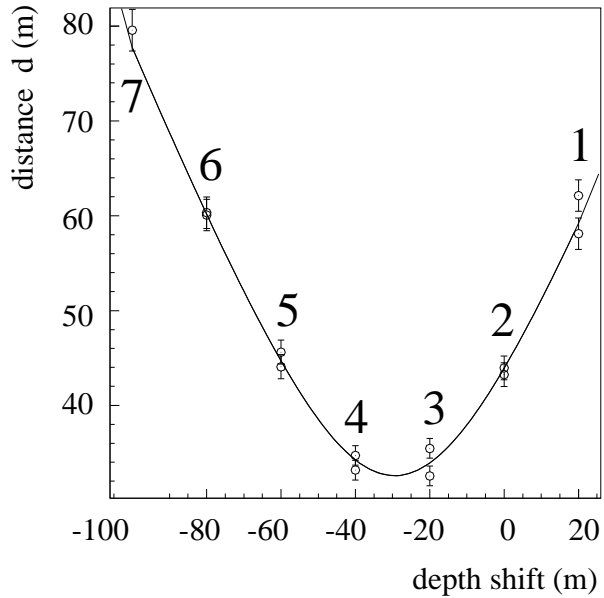,height=8cm}}
\caption{\small {Fit of the distance $d(z_i)$ versus depth-shift 
$z_i - z_2$ between OMs at string 4 and
a laser emitter at string 2. String distance $D$ and 
depth shift $z_0 - z_2$ are given by the minimum
of the parabola.}}
\label{yag-string2}
\end{center}
\end{figure}

Given the limited number of measured emitter-OM combinations available for 
AMANDA-B4, it
would have been impossible to keep the coordinates of 
each OM  as free
parameters in a global position fit. 
Therefore, all strings were
assumed to be straight and parallel and the OMs to be at a fixed
vertical distance (20\,m) relative to each other. For each emitter
covering enough OMs, a graph of the distance $d(z_i)$
between source and OM $i$ versus depth $z_i$  can be drawn 
(see fig.~\ref{pos_princip}). The inter-string distance $D$ and 
emitter depth $z_0$ with respect to the $z_i$
can be estimated from this graph by fitting 
(fig.~\ref{yag-string2}):

\begin{equation}
\label{eq:distance_function}
d(z_i) = \sqrt{D^2+(z_i-z_0)^2}.
\end{equation}

The residuals from all fits to the 1996 AMANDA-B4 data
have a standard deviation of 2 m.

In 1996--1997, six more strings were added on the outside 
of the B4 detector, and a new position calibration performed. The 
increased statistics and possibilities of new cross-checks and
constraints  enabled correction of the existing geometry with
an uncertainty of 1\,m in the horizontal plane and 0.5-1.0\,m in
depth.

\medskip

{\large \it Drill data}

The geometry of the array is surveyed in an
independent way by monitoring the position of the
drill-head while it is going down each hole.
The data were recorded by the drill instrumentation at each 10 cm step,
recording the path-length, the value of the Earth's magnetic field as
measured by a flux magnetometer and the angles (bank and elevation)
given by perpendicular pendulums. This information can then be used to
reconstruct the hole profiles. The results found are compatible
with the laser measurements within 1-2 m in the horizontal plane.
The advantage of this method is that it yields positions relative to
the surface, i.e. in a global reference frame. It  also takes into
account tilts in the strings. However, it does not yield the depth
locations of the OMs. The absolute depths of the strings were given by
pressure sensors deployed with the OMs.

%\newpage

\section{Simulation and Reconstruction of Muons \label{simureco}}

\subsection{Simulation \label{simulation}}

%Upgoing and downgoing muons are simulated  by
%separate programs. They are then used as the input for
%a simulation of AMANDA itself, producing events in the same 
%format as  the experimental data. 

Downgoing muons are  generated by  full atmospheric shower programs
which simulate the production of muons by isotropic primary
protons \cite{Boziev}  or protons and nuclei \cite{Hemas} with
energies up to 1 PeV. The muons are propagated
down to a plane close to the detector.  
Upgoing muons  are generated from 
atmospheric neutrinos, using the flux parameterization
given in \cite{Volkova}, from neutralinos annihilating in the
center of the Earth, using the 
flux calculations of \cite{WIMP}, 
and from  point sources, using arbitrary
energy distributions and source angles; they may start anywhere within
the fiducial volume (which increases with increasing neutrino
energy due to the muon range) and are propagated simulating the
full stochastic energy loss according to \cite{Lohmann}.

\pagebreak

It would be computationally impractical to generate
and follow the path of each of the multiply scattered Cherenkov
photons produced by  muons and
secondary cascades for every simulated event. Therefore, this step is
accomplished by doing the photon propagation only once by a
separate MC program and storing
the results in large multidimensional tables. The tables give the
distribution of the mean number of photoelectrons expected and of
the time delay distribution, as a
function of the position and the orientation of a PMT 
relative to the muon track. They include the effects of the
wavelength dependent quantum efficiency, the transmission
coefficients of glass spheres and optical gel, and the
absorption and scattering properties of the ice. Once the tables
are compiled, events can be simulated quickly by locating the
PMT relative to any input particle and looking up the expected
number and time distribution of photoelectrons in the tables\footnote{This 
method assumes that ice is isotropic and
and homogeneous which is reasonable in
a first approximation: firstly, since
the variations of the original ice with depth have been 
measured to be smaller than $\pm 25\%$, secondly, since the freshly 
frozen ice in the holes occupies only a small volume of the array.}. 
The known characteristics of the AMANDA PMTs, the
measured pulse shapes, pulse heights and delays after
signal propagation along the cables, and the effect of  electronics 
are then used to generate amplitude and time information
\cite{Stephan}. 

\subsection{Reconstruction \label{reco}}

The reconstruction procedure for a muon track consists of
five steps:

\begin{enumerate}
\vspace{-2mm}
\item Rejection of noise hits, i.e. hits which have either a
very small ADC value or which are isolated in time with respect
to the trigger time or with respect to the nearest hit OM.
\vspace{-1mm}
\item  A line approximation following \cite{Stenger} which yields a 
point on the track, $\vec{r}$, and a velocity 
$\vec{v}$:
\vspace{-1mm}
$$
\vec{r} = \langle r_i \rangle - \vec{v} \cdot \langle t_i \rangle  
\hspace{2cm} \vec{v} =  \frac{\langle \vec{r}_i t_i \rangle - 
\langle \vec{r}_i \rangle  \langle t_i \rangle}
       {\langle t_i^2 \rangle - \langle t_i \rangle^2}.
$$
\vspace{-1mm}
with $\vec{r_i}$ and $t_i$ being the coordinate vector and response time
of the $i$-th PMT.
\vspace{-1mm}  
\item A likelihood fit based on the measured times which takes 
         the the track parameters obtained from the line
         fit as start values.  
         This "time fit" yields angles and coordinates
        of the track as well as a likelihood ${\cal L}_{time}$.
\vspace{-1mm}
\item A likelihood fit using the fitted track parameters from the time fit
         and varying the light emission per unit length until the probabilities
         of the hit PMTs to be hit and non-hit PMTs to be not hit are
         maximized. This fit does not vary the direction of the track
         but yields a likelihood ${\cal L}_{hit}$
         with can be used as a quality parameter.
\vspace{-1mm}         
\item A quality analysis, i.e. application of cuts in order to reject badly 
          reconstructed events.
\vspace{-1mm}
\end{enumerate}

Steps 3 and 5 are outlined in the following two subsections. 
       
\subsection{Time Fit}

\begin{figure}[htbp]
\begin{center} 
\mbox{\epsfig{file=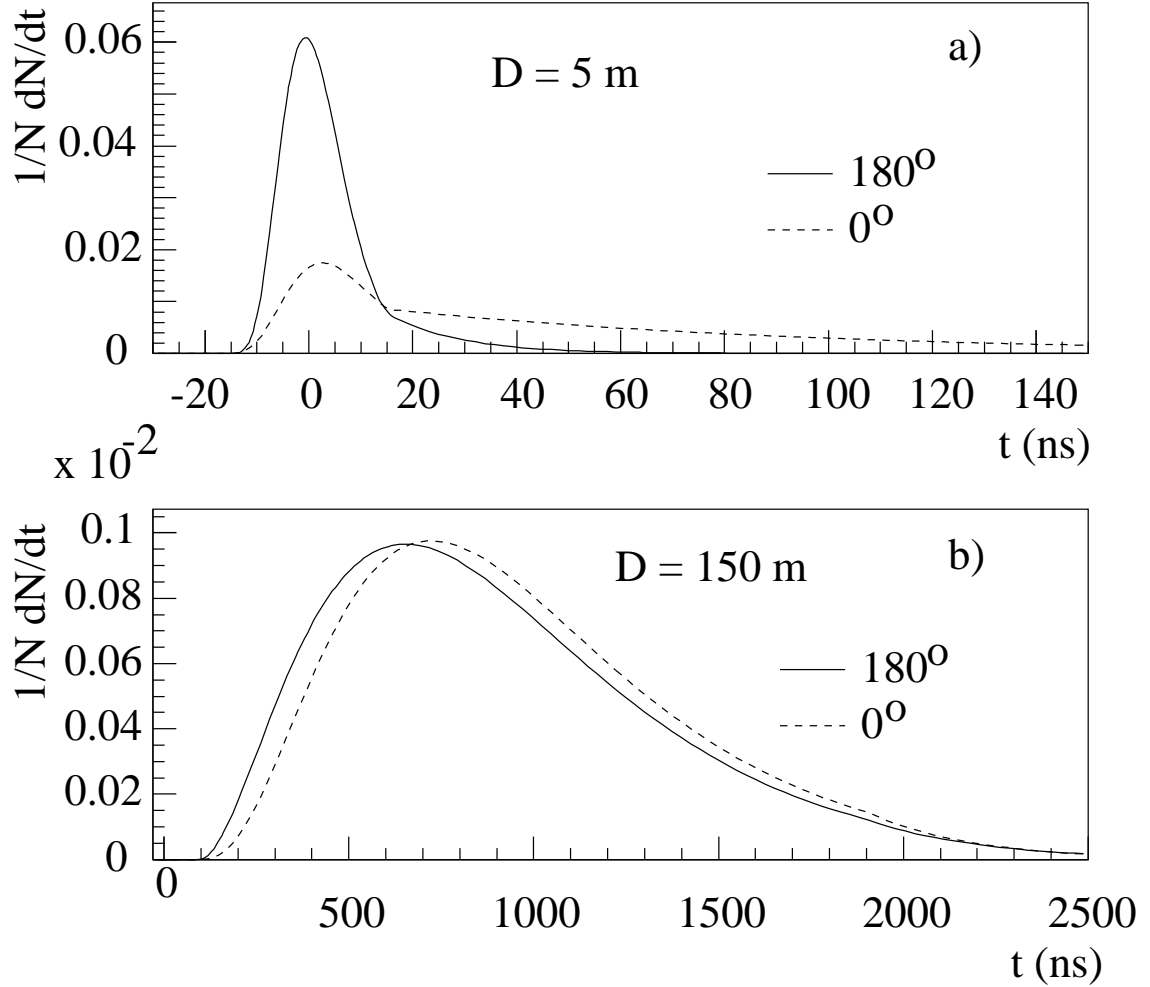,height=13cm}} 
\caption[2]{\small
Delay-time distributions for modules facing (full curves) and 
back-facing (dashed curves) a muon track. Data are shown for
muon tracks with impact 
parameters of 5 meters (a) and 150 meters (b).
}
\label{adam}
\end{center}
\end{figure}

In an ideal medium without scattering, one would reconstruct
the path of  minimum ionizing muons most efficiently 
by a $\chi^2$ minimization process.
Due to  scattering in ice, the distribution of arrival times
of photoelectrons seen by a PMT is not Gaussian
but has a long tail at the high side -- see fig.~\ref{adam}. 
%Fig.~\ref{cer_tres}.

To cope with the non-Gaussian timing distributions
we used a likelihood analysis. In this approach, a
normalized probability distribution function $p_i(t)$  gives the 
probability of a certain time delay $t$ for a given hit $i$
with respect to straightly propagating photons. 
This probability function is derived from the MC simulations
based on the photon propagation tables introduced in section~\ref{simulation}.
The probability 
depends on the distance and the orientation of the PMT with respect
to the muon track. 
By varying the  track parameters the logarithm of a 
likelihood function ${\cal L}$  is maximized.

$$
\log ({\cal L}) = \log \left ( \prod_{\mbox{all hits}} p_i \right )
= \sum_{\mbox{all hits}} \log ( p_i ) 
$$

In order to be used in the iteration process, the time delays  as obtained
from the separate photon propagation Monte-Carlo have to be parameterized
by an analytic formula.
The parameterization of the
propagation model itself is
extended to allow for timing errors of PMTs 
and electronics as well as
the probability of noise hits at random times.
The AMANDA collaboration has developed two independent reconstruction
programs, which are based on different parameterizations of
the photon propagation and different minimization methods 
\cite{Bouchta,icrc_reco,wieb2}.
The comparison of these algorithms and the use of 
different  optical models show that the results of
both methods are in good agreement with each other
and do not depend on a fine-tuning 
of the assumed optical parameters.
Fig.~\ref{adam} shows the result of the parameterization of the time
delay for two distances and for two angles between the PMT axis
and the muon direction.
At a distance of 5\,m and a PMT facing toward the muon
track, the delay curve is dominated by the time jitter of the
PMT. However, if the PMT looks in the opposite direction,
the contribution of scattered photons yields a long
tail towards large delays. At distances as large as 150\,m,
distributions for both directions of the PMT are close to each
other since all photons reaching the PMT are multiply scattered.

The  parameterization  used for most
of the results presented in this paper is
a Gamma distribution modified with an absorption term \cite{Pandel}

$$ 
p(d,t) = N \cdot { \tau^{-(d/\lambda)} \cdot t^{(d/\lambda -1)} \over \Gamma
(d/\lambda ) }
\cdot e^{- t/\tau + c_w \cdot t/X_0 + d/X_0 } ~,
$$

with the distance $r$ between OM and muon track,
the  scaled distance $d \approx 0.8 / \sin (\theta_c ) \cdot  r  $,
 the absorption length $X_0 $ and only two parameters 
$\tau \approx 700  $\,ns
and $ \lambda \approx 50$\,m. 

The second approach uses an F-distribution with an exponential
tail for large time-delays, which results in a comparable accuracy
\cite{Bouchta}.

\subsection{Quality Analysis}

%The reconstruction proceeds in a multi-step analysis:
% After rejection  of isolated noise hits, 
% a first track approximation is  calculated from a line-fit
% \cite{Stenger}. 
% This is followed by one or more subsequent likelihood reconstructions.
% In the last step a quality analysis is performed in order to identify
% badly reconstructed tracks. 

Quality criteria are applied in order to
select events which are "well" reconstructed.  
The criteria define cuts on topological event parameters
and observables derived from the 
reconstruction.
Below we list those used in the following:

\begin{itemize}

\item 
Speed $|\vec{v}|$ of the line fit. Values close to the speed of light
indicate a reasonable pattern of the measured times, values smaller
than 0.1 m/nsec indicate an obscure time pattern.

\item
"Time" likelihood per hit PMT $\log({\cal L}_{time})/N_{hit}$. 

\item
Summed hit probability for all hit PMTs $\sum P_{hit}$. 
 
\item 
"Hit" likelihood normalized to all working channels, 
$\log({\cal L}_{hit})/N_{all}$.

The latter two parameters are good indicators of whether the
location of the fitted track, which relies exclusively on the
time information, is compatible with the location of the hits
and non-hits within the detector.

\item 
Number of direct hits, $N_{dir}$, which is defined to be
the number of hits with time residuals $t_i\mbox{(measured)} - 
t_i\mbox{(fit)}$
smaller than a certain cut value. We use cut values of 15\,nsec,
25\,nsec and 75\,nsec, and denote the corresponding
parameters as $N_{dir}$(15), $N_{dir}$(25) and
$N_{dir}$(75), respectively. Increasing the time
window leads to higher cut values in $N_{dir}$ but allows
a finer gradation  of the cut.

Events with more than a certain minimum number of direct
hits (i.e. only slightly delayed photons) are likely to be
well reconstructed. This cut turned
out to be the most powerful cut of all \cite{icrc_reco}.

\item
The projected length of direct hits onto the reconstructed 
track, $L_{dir}$.
A cut in this parameter rejects events with a small lever arm.

%\item
%Distance to the detector center, $L_{center}$.

\item
Vertical coordinate of the center of gravity, $z_{COG}$.
Cuts on this parameter are used to reject events close
to the borders of the array. Very distant tracks are not likely to be
well reconstructed.
\end{itemize}

\begin{figure}[htbp]
\centering
%\mbox{\epsfig{file=myserap3.eps,height=10cm}}
\mbox{\epsfig{file=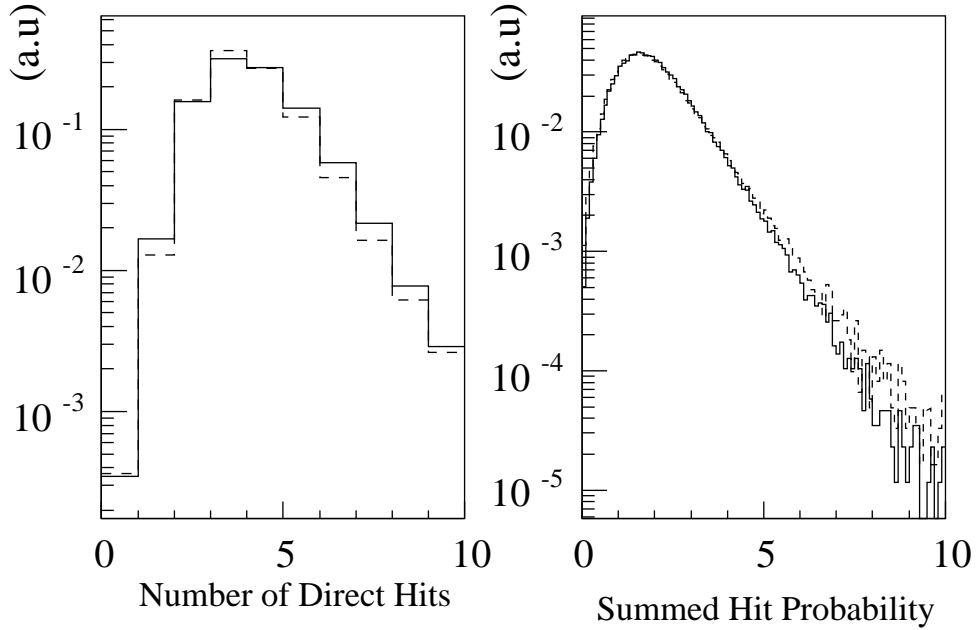,width=13cm}}
%~ \hfill (A)  \hfill (B)  \hfill
\caption[2]{\small
Distributions of two reconstructed
event observables for MC down-going muon events 
(dashed lines) and from
experimental data (full lines). 
 {\it Left:} Number of direct hits,
$N_{dir}$(15); 
{\it Right:} summed hit probability, $\sum{P_{hit}}$,
}
\label{serap3}
\end{figure}

Fig.~\ref{serap3}  shows the distribution of
two of these observables, the number of direct hits within
15\,nsec, $N_{dir}$(15),
and the summed hit probability $\sum{P_{hit}}$ of all hit channels.
It demonstrates the
good agreement between results from MC and experiment.  

\begin{figure}[htbp]
\centering
\mbox{\epsfig{file=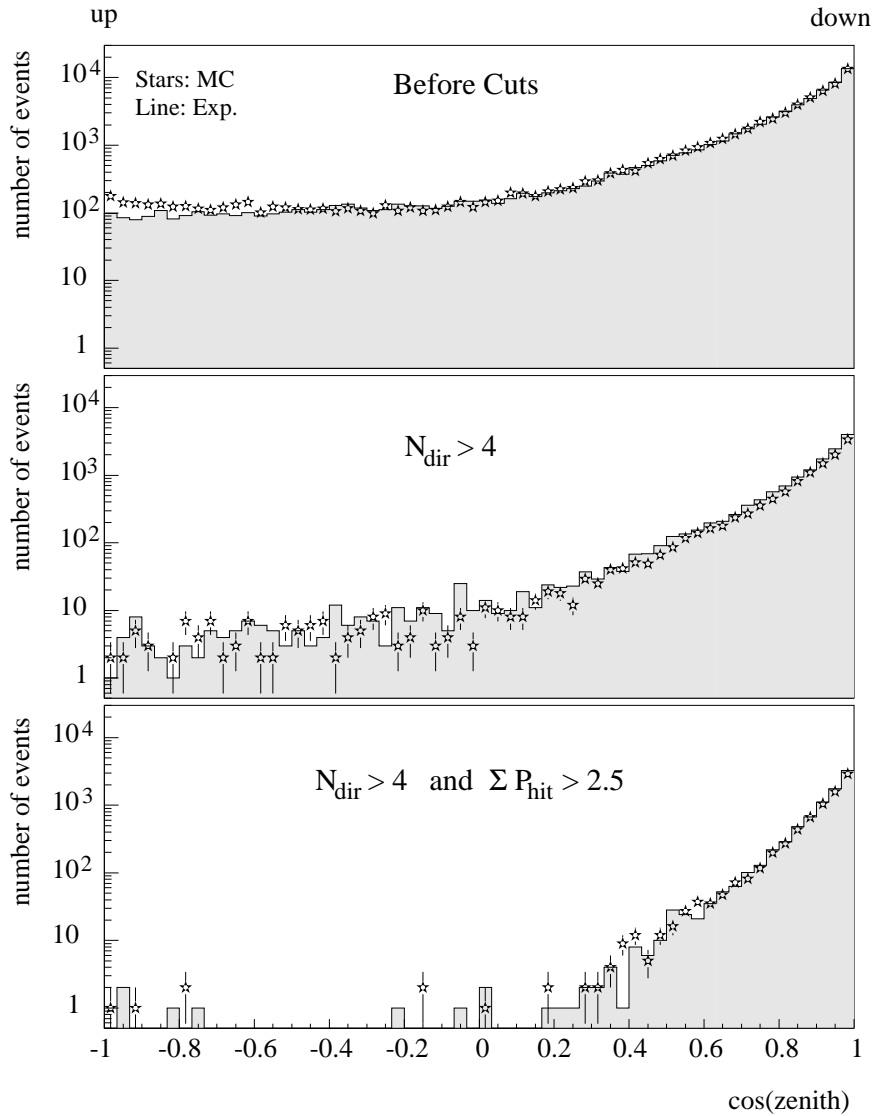,width=11.5cm}}
%\mbox{\epsfig{file=all_ang_4str.vrt.eps,height=17cm}}
\caption[2]{ \small 
Reconstructed zenith angle distributions of experimental
data (line) and downward muon MC events (points)
after a stepwise application of quality cuts.
}
\label{serap4}
\end{figure}

Fig.~\ref{serap4} demonstrates the effect of cuts on the 
number of
direct hits and the summed hit probability
on the reconstructed
angular distribution of experimental
data and the MC sample.
The cuts are $N_{direct}\mbox{(15)} \ge 5$
and $\sum{P_{hit}} \ge 2.5$.
Both samples are dominantly due to down-going atmospheric muons.
Despite that, a small but similar fraction of events is 
falsely reconstructed
as up-going events. After application of
the above quality criteria the tail below the horizon almost
disappears. Note that not
only the shapes but also the absolute passing rates
on all cut levels
are in good agreement between data and Monte Carlo.
The angular mismatch
between the reconstructed muon angle and the original angle
used in the MC simulation after both cuts 
is 5.5 degrees. We note that this value strongly
depends on the particular set of cuts, the minimum
acceptable passing rate, the incident angle of the muon,
and the range of muons stopping in the array.

\section{SPASE coincidences \label{spase}}

\begin{figure}[htbp]
\centering
\mbox{\epsfig{file=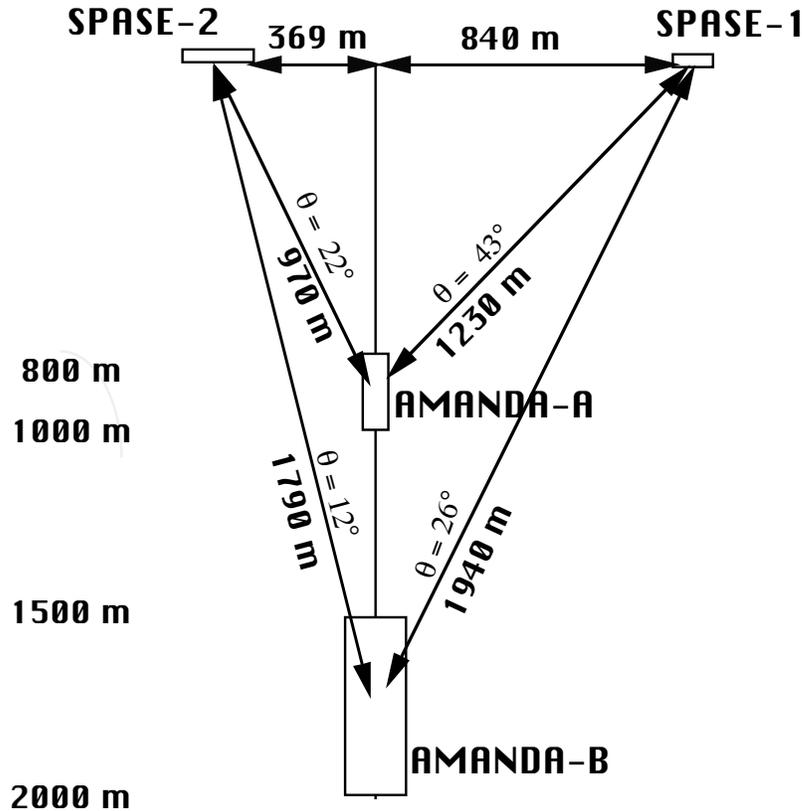,height=11cm}}
%\mbox{\epsfig{file=sideview.ps,height=8cm}}
\caption [10]
{\small Side view of the two SPASE arrays relative to 
AMANDA-A and AMANDA-B.
}
\label{Spase1}
\end{figure}

AMANDA is unique in that it can be
calibrated by muons with
known zenith and azimuth angles which are tagged by 
air shower detectors at the surface. AMANDA-B4 has been running in
coincidence with the two SPASE (South Pole Air Shower Experiment)
arrays, SPASE-1 \cite{Beaman93}
and SPASE-2 \cite{Gais95}. SPASE-1 was located 840 m
from the center of the AMANDA array projected to the
surface, whereas SPASE-2 is located
370\,m away (see fig.\ref{Spase1}). 
The scintillation detectors of SPASE-2 are complemented by
an array of air Cherenkov detectors \cite{Spase2,Vulcan}. 
The primary goal of these
devices is the investigation of the chemical composition of
primary cosmic rays in the region of the "knee" \cite{Miller97}.
Another detector, the gamma imaging telescope GASP, 
is also operated in coincidence with AMANDA.

In this section, we summarize calibration results 
obtained from the coincident operation
of AMANDA and SPASE-2. SPASE-2 consists of 30 scintillator stations
of 0.8\,m$^2$ on a 30\,m triangular grid. The area of the array is 
$1.6 \cdot 10^4$\,m$^2$, and it has been running since January 1996.
For each air shower, the direction,
core location, shower size and GPS time are determined. 
Showers
with sufficient energy to trigger SPASE-2  ($\approx$ 100\,TeV)
yield on average 1.2 muons  penetrating
to the depth of  AMANDA-B. 
On every SPASE-1 or SPASE-2 trigger, a signal is sent
to trigger AMANDA.
The GPS times of the separate events
are compared offline to match coincident events.

\begin{figure}[htbp]
\centering
\mbox{\epsfig{file=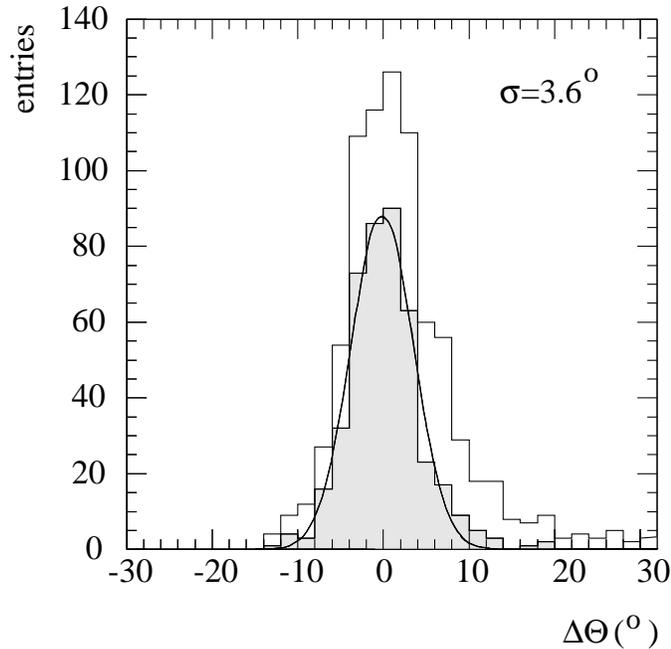,height=8.5cm}}
\caption [11]{\small
Mismatch between zenith angles determined in AMANDA-B4 and SPASE-2.
}
\label{Spase2}
\end{figure}

A one-week sample 
of these events has been analyzed in order to compare
the directions of muons determined by AMANDA-B4 to those
of the showers measured by SPASE-2. A histogram of the zenith mismatch 
angle between SPASE-2 and AMANDA-B4 
is  shown in fig.\ref{Spase2}. 
The selected events are required to 
have $\ge$8 hits along 3 strings
and to yield a track which is closer than 150\,m to the
air shower axis measured by SPASE-2 (upper histogram).
The hatched histogram shows the distribution of the zenith
mismatch angle after application of the following quality cuts:

\begin{itemize}

\item  likelihood $\log({\cal L}_{time})/N_{hit} > $ -12,
\item  more than four hits with residuals smaller than 75 nsec
($N_{dir}\mbox{(75)} > 4$), 
\item  length of the projection
of OMs with direct hits to the track larger than 50 meters
($L_{dir}\mbox{(75)} > 50$\,m). 

\end{itemize}

428 of the originally 
840 selected events pass these
quality cuts. The Gaussian fit has a mean of
$(0.14 \pm 0.19)$ degrees and a width of $\sigma = (3.6 \pm 0.17)$ 
degrees.  This is nearly 2 degrees better than the resolution
obtained in the previous section for {\it all} downward
muons and for a different set of cuts. MC yields a resolution of about
4 degrees.

The small mean implies that there is little 
systematic error in zenith angle
reconstruction.
The SPASE-2 pointing accuracy,
which contributes to the average mismatch,
depends on zenith angle and shower size.
For most of the coincidence events, the SPASE-2 pointing
resolution, defined as the angular distance within which
63$\%$ of events are contained,
is between 1$^\circ$ and 2$^\circ$ \cite{Spase2, Vulcan}.

\section{Intensity-vs-Depth Relation for Atmospheric Muons \label{depth}}

\subsection{Angular Dependence of the Muon Flux}

\begin{figure}[htbp]
\centering
\mbox{\epsfig{file=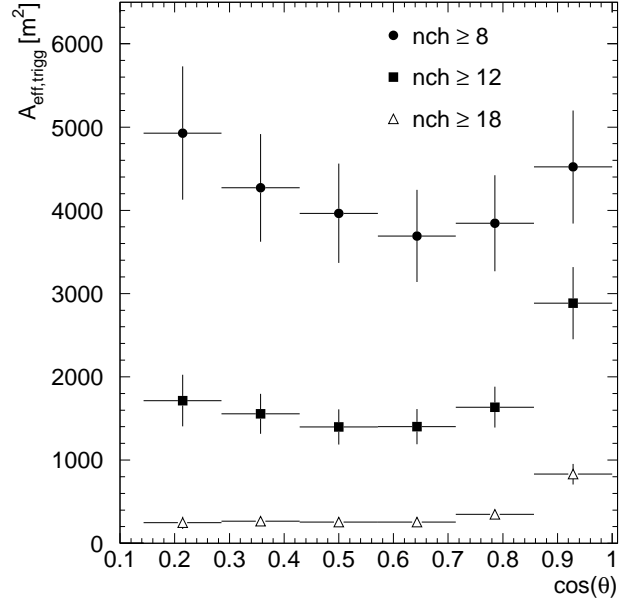,height=9cm}}
%\mbox{\epsfig{file=aefft.eps,height=7cm}}
\caption[2]{\small
Effective trigger area of AMANDA-B4 as a function
of zenith angle, for 3 different majority criteria.
}
\label{aeff}
\end{figure}

In section~\ref{simureco}, 
the muon angular distribution was shown as a function
of various cuts in order to demonstrate the agreement between
experimental data and MC simulations. 
In this section, we calculate the muon intensity $I$
as a function of the zenith angle $\theta$.
$I(\theta_{\mu})$ is given by

\begin{equation}\label{fluxform1}
   I(\theta_{\mu})=   \frac{S_{dead}}{T \cdot \Delta \Omega}\, 
%   \sum_{\theta_{rec}}
% zaehler   
   \frac{ 
%  N_{\mu}(\theta_{rec})\, 
   N_{\mu}(\theta)\,
%   f(\theta_{rec},\theta_{\mu}) 
   \cdot m(\theta_{\mu})}
% nenner
   {{\epsilon_{rec}(\theta_{\mu})} \cdot
   {A_{eff}}(\mbox{cut},\theta_{\mu})}
\end{equation}
where
\vspace{-2mm}
\begin{itemize}

\item  $N_{\mu}(\theta$) is the number of muons assigned by the
       analysis to a zenith angle interval centered around
       $\cos \theta_{\mu}$. For the analysis presented in this
       section, we start from the angular
       distribution 
       $N_{\mu}(\theta_{rec})$ obtained from
       the reconstruction, without applying cuts. This distribution
       is strongly smeared (see fig.\,\ref{serap4}, top).
       We have calculated the elements of the parent
       angular distribution  $N_{\mu}(\theta)$ from the
       reconstructed distribution  $N_{\mu}(\theta_{rec})$
       using a regularized deconvolution procedure
       \cite{Decon,Blobel}.

\item  $T$ is the run time. We used the data from June 24, 1996,
       with $T=22.03$ hours, and 9.86\,$\cdot \,10^5$ events triggering
       AMANDA-B4.

\item  $S_{dead}$ corrects for the dead time of the data
       acquisition  system. 
       This factor was determined  from the
       time difference distribution of subsequent events.
       The dead-time losses for the two
       runs used in this analysis are 12\%, i.e. $S_{dead}=1/0.88 = 1.14$.

\item  $\Delta\Omega$ is the solid angle covered by the 
       corresponding $\cos{\theta_{\mu}}$ interval.

\item $A_{eff}(\mbox{cut},\theta_{\mu})$ is the effective area, after
      the application of a multiplicity trigger,
      for a given cut at zenith angle $\theta_{\mu}$. The effective
      area is shown in fig.~\ref{aeff} as a function of the zenith
      angle and for different cuts on the number of hit OMs.

\item $\epsilon_{rec}(\theta_{\mu}$) is the reconstruction efficiency for
      zenith angle $\theta_{\mu}$ which ranges between 0.82 at 
      $\cos \theta = 1.0$ and 0.75 at $\cos \theta = 0.2$. 

\item  $m(\theta_{\mu})$ is the mean muon multiplicity at angle 
       $\theta_{\mu}$ at the "trigger depth". 
       The trigger depth
       $h_{eff}$ was defined as 
%the average 
depth of
$\overline{z_{OM}}$, the center of gravity in the vertical coordinate $z$ 
of all hit OMs.
%, and $z_{OM,min}$ , the $z$ coordinate of the lowest hit OM.  
The average $h_{eff}$ depends on the angle. 
It is highest  for $\cos\theta$ between 0.4 and 0.8 
(about 30 m below the detector center)
and falls toward the vertical (at maximum 80\,m below the center).
       The mean muon multiplicity is about 1.2 for vertical tracks and
       decreases towards the horizon.
       Since the generator used in this analysis \cite{Boziev} simulates
       only proton induced showers, this value  is an
       underestimation by about 10\%.  

\end{itemize}

\begin{figure}[htbp]
\centering
\mbox{\epsfig{file=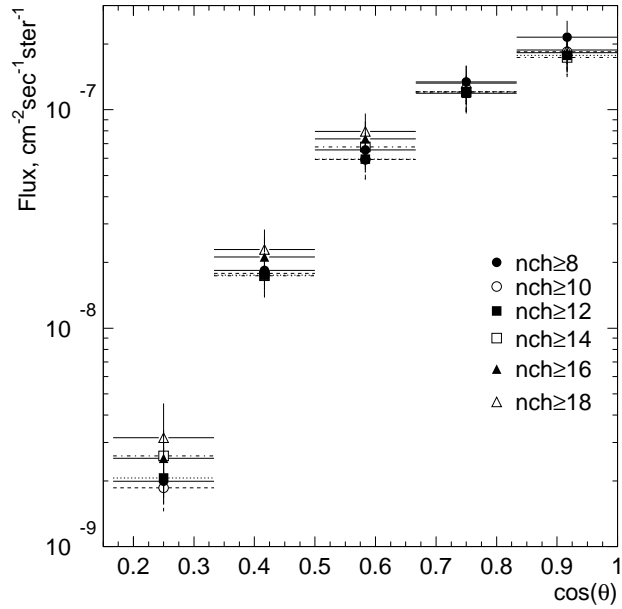,height=9cm}}
%\mbox{\epsfig{file=angularflux.eps,height=7cm}}
\caption[2]{\small
Angular distribution of the downward going muon flux, 
$I(\theta_{\mu})$, as obtained from eq.(3). 
}
\label{fluxtheta}
\end{figure}

Fig.~\ref{fluxtheta} shows the angular distribution of the
flux of downgoing muons, $I(\theta_{\mu})$, as obtained from
eq.3. In order to illustrate
the stability of the method with respect to cuts biasing
the measured angular distribution, the flux is shown
for samples defined by different majority triggers
($N_{hit} >$ 8,\,10,\,12,\,14,\,16,\,18). Apart from the point closest
to the horizon which is not only most strongly biased but also has 
the lowest statistics, deviations are within  25\%.
For further studies we use the sample with $N_{hit} \ge 16$.

\subsection{Transformation of  Angular Flux to 
Vertical Intensity as a Function of Depth}

The measured flux $I(\theta)$ can be transformed into a
vertical flux $I(\theta=0,h)$, where $h$ is the ice thickness
seen under an angle $\theta$:

\begin{equation}
I(\theta=0,h)=I(\theta) \cdot \cos(\theta) \cdot c_{corr}
\end{equation}

The $\cos \theta$-conversion correcting for the
sec($\theta$) behavior of the muon flux is valid
for angles up to 60$^o$ \cite{Gaisser}.
The term  $c_{corr}$ taken from \cite{Lipari} corrects
for larger angles and lies between 0.8 and 1.0 for the angular 
and energy ranges considered here.

\begin{figure}[htbp]
\centering
\mbox{\epsfig{file=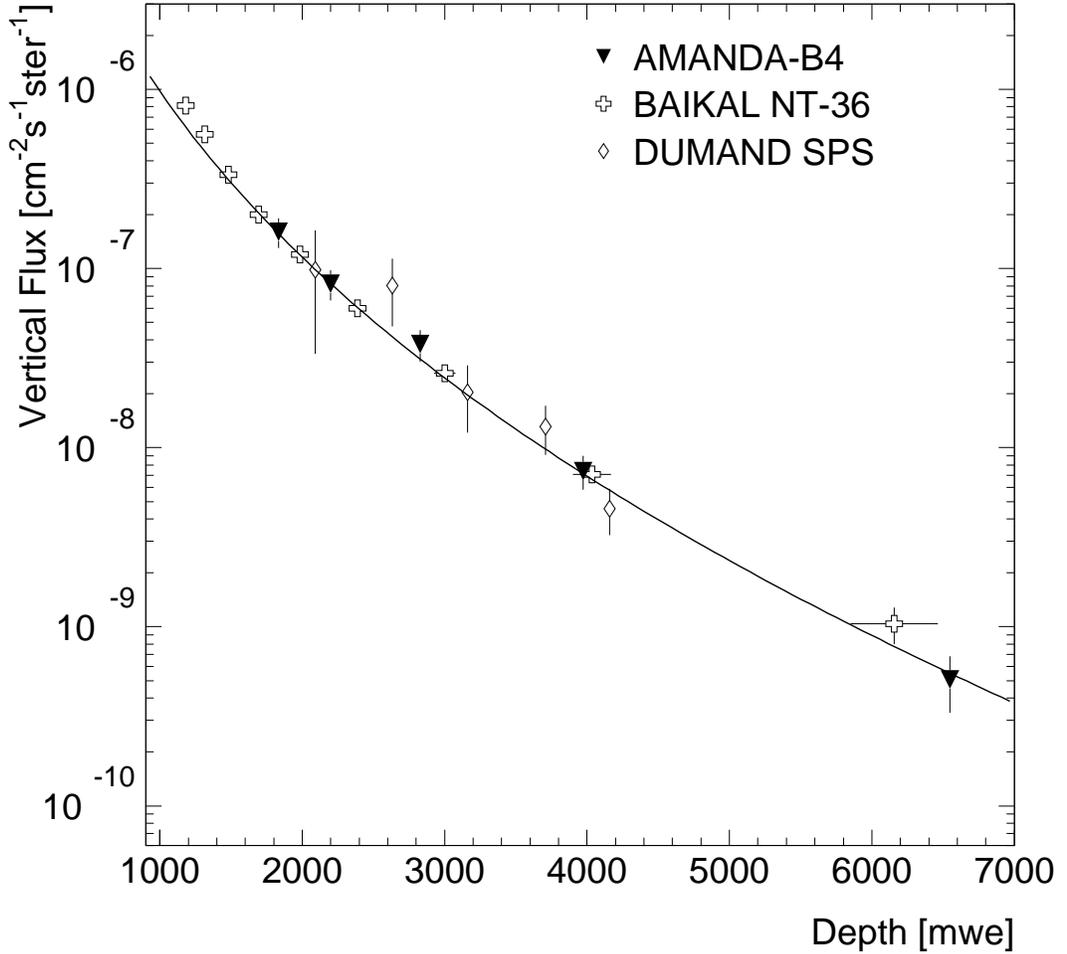,height=14cm}}
\caption[2]{\small
Vertical intensity versus depth for AMANDA,
BAIKAL and DUMAND.
The solid line gives the prediction of
\cite{Bugaev} which coincides with the curves obtained from
the parameterizations (5) and (6).
}
\label{ABD}
\end{figure}

The vertical intensities obtained in this way are plotted in
fig.~\ref{ABD} and compared to 
%the expectation from \cite{Bugaev} (full line). In fig.~\ref{ABD} 
%we compare our results to 
the depth-intensity data published by DUMAND \cite{SPS} and Baikal
\cite{Baikal}, and to the prediction by Bugaev et al. \cite{Bugaev}.
One observes satisfying agreement of all experiments with the
prediction.
% as well as between each other.

We also fitted our data to
a parameterization taken from \cite{Allkover,Rhode}:

\begin{equation}
I(h)=I_{0}\cdot E_{crit}^{-\gamma}=I_{0}\cdot 
\left(
\frac{a}{b_{eff}}\cdot \left[ e^{(b_{eff} \cdot h)}-1 \right]
\right)^{-\gamma}
\end{equation}

$E_{crit}$ is the minimum energy necessary to reach the depth $h$.
It is obtained from the parameterization \cite{Gaisser}
$ dE/dx = a + b \cdot E_{\mu} $ where $a \approx $ 2 MeV/(g$\cdot$cm$^{-2}$)   
denotes the continuous energy loss due to ionization, and 
$b(E_{\mu})$ is proportional to the
stochastic energy loss due to pair production,
bremsstrahlung and nuclear cascades. From this parameterization
one obtains $E_{crit} = a/b \cdot [\mbox{exp}(b \cdot h) - 1]$.
$I_{0}$ is the normalization parameter
%($\mbox{cm}^{-2}\,\mbox{s}^{-1}\,\mbox{ster}^{-1}$),
%a $\approx$ 2\,MeV g/cm$^2$ is the ionization energy loss
and $\gamma \approx$ 2.78 \cite{Rhode} the spectral index.
%The parameter $b$ gives the energy loss due to
%stochastical processes.
We approximate $b(E_{\mu})$ by an energy independent parameter
$b_{eff}$. Fitted to equation (5), our data  for the
vertical intensity result in the following values for
$I_{0}$ and $b_{eff}$:

\begin{table}[H]
\centering
\begin{math}
\begin{array}[H]{l}
   I_{0}=(5.04 \pm 0.13)
   \, \mbox{cm}^{-2}\mbox{s}^{-1}\mbox{ster}^{-1}\\
   b_{eff}=(2.94 \pm 0.09)\,\cdot \, 10^{-6}\,\mbox{g}^{-1}\, \mbox{cm}^2.\\
\end{array}
\end{math}
\end{table}

This compares to $I_{0}=(5.01 \pm 0.01)
   \, \mbox{cm}^{-2}\mbox{s}^{-1}\mbox{ster}^{-1}$ and
   $b_{eff}=(3.08 \pm 0.06)\,10^{-6}\,\mbox{g}^{-1}\, \mbox{cm}^2$
obtained for $N_{hit} \ge 8$, showing that the result is rather
insensitive to the actual cut condition.

For the purpose of completeness we give also the results for
the more usual parameterization

\begin{equation}\label{macrofunc}
   I(h,\theta_{\mu}=0)=a_{\mu}\left(\frac{\lambda}{h}\right)^\alpha
   \,\mbox{e}^{-h/\lambda}
\end{equation}

where $\alpha$ is set to 0 \cite{CWI}, to 2 \cite{Frejus} or 
is a free parameter \cite{Macro}. The purely
exponential dependence ($\alpha = 0$) clearly does not describe the
data at depths smaller than 4-5 km. Leaving all parameters
free \cite{Macro}, one obtains $a_{\mu}= (0.89 \pm 0.30) \cdot 10^{-6}
   \, \mbox{cm}^{-2}\mbox{s}^{-1}\mbox{ster}^{-1}$,
$\lambda = (1453 \pm 612)$\,g\,cm$^{-2}$, and $\alpha = 2.0 \pm 0.25$,
being also in agreement with
$\alpha$ fixed as in \cite{Frejus}.

%\newpage
 
\section{Search for Upward Going Muons \label{upward}}

AMANDA-B4 was not intended to be a 
full-fledged neutrino detector, 
but instead a device which demonstrates the 
feasibility of muon track
reconstruction in Antarctic ice.  
The  limited number of optical modules and
the small lever arms in all but the vertical direction 
complicate the rejection of fake events. 
In this section we demonstrate
that in spite of that
the separation of a few upward
muon candidates was possible.

We present the results of two independent analyses.
One uses the approximation of the likelihood function
by a F-function with an exponential tail \cite{Bouchta}, the
other the approximation by a Gamma function with an absorption term
\cite{wieb2} (see section 6.3).

Both analyses apply separation criteria which are obtained from a 
stepwise tightening of cuts on different parameters, in a way
which improves the signal-to-fake ratio given by the
MC samples. Since the MC generated samples of downward-going muons
(a few million events) run out of statistics after a reduction
factor of about $10^6$, further tightening of cuts is
performed without background-MC control until the 
experimental sample reaches
the same  magnitude as the MC predicted signal.

For both analyses,
the full experimental data set of 1996, 
starting with Feb.19th and ending with Nov.5th, was processed.
It consists of $3.5 \cdot 10^8$ events.

\bigskip

{\large \it Analysis 1}

In a first step, a fast pre-filter
reduced this sample to a more manageable size.
It consists of a number of cuts on quickly computable variables
which either correlate with the muon angle, or which to a certain
degree distinguish single muons from the downgoing multi-muon
background events like, e.g. a cut on the zenith angle from
a line fit \cite{Stenger}, cuts on time differences between
OMs at different vertical positions, and topological cuts requesting
a minimum vertical elongation of the event.

These cuts reduce the size of the experimental data sample to 5.2\%,
the simulated atmospheric muons to 4.8\% and simulated up-going
events to 49.8\%.

%Due to CPU limitations, we did not attempt to simulate
%a six-month sample of downgoing muons.
%Therefore only 
Simulated up-going events and experimental data
%are considered in the following. Both 
have been reduced by further cuts:

\begin{itemize}

\item At least 2 strings have to be hit
      %The hits are from $\ge$ 2 strings 
      (this condition relaxes the
      standard condition  "$\ge$3 strings" and increases the 
      effective area in the vertical direction).

\item The events were reconstructed below horizon, i.e. $\theta >$ 90$^o$.

\item $\log({\cal L}_{time})/N_{hit} > -6$. 

\item $\alpha \ge$ 0.15 m/nsec, where $\alpha$ is obtained
from a fit to $z_i = \alpha \cdot t_i + \beta$ and $z_i,\,t_i$
being the $z$\,coordinates and times of the hit OMs. 

\end{itemize}

\begin{figure}[htbp]
\centering
\mbox{\epsfig{file=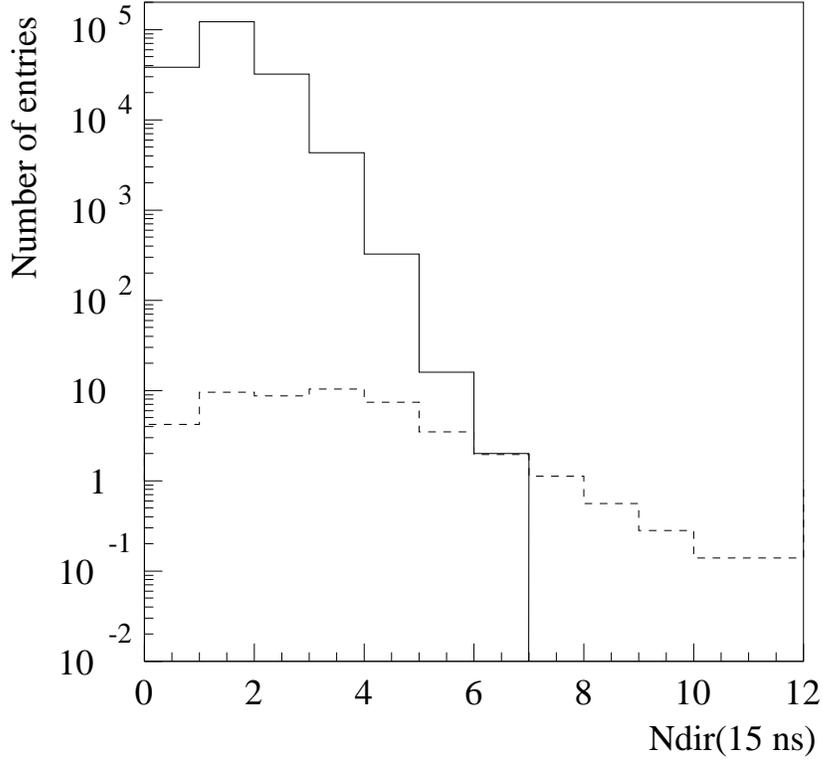,height=10cm}}
\caption[77]{\small
Number of events surviving pre-filter and additional cuts as a
function of $N_{dir}$(15). 
Solid line: 6-month experimental
data. dashed line: 6-month expectation from atmospheric neutrinos.
}
\label{ndirect}
\end{figure}

\begin{figure}[htbp]
\centering
\mbox{\epsfig{file=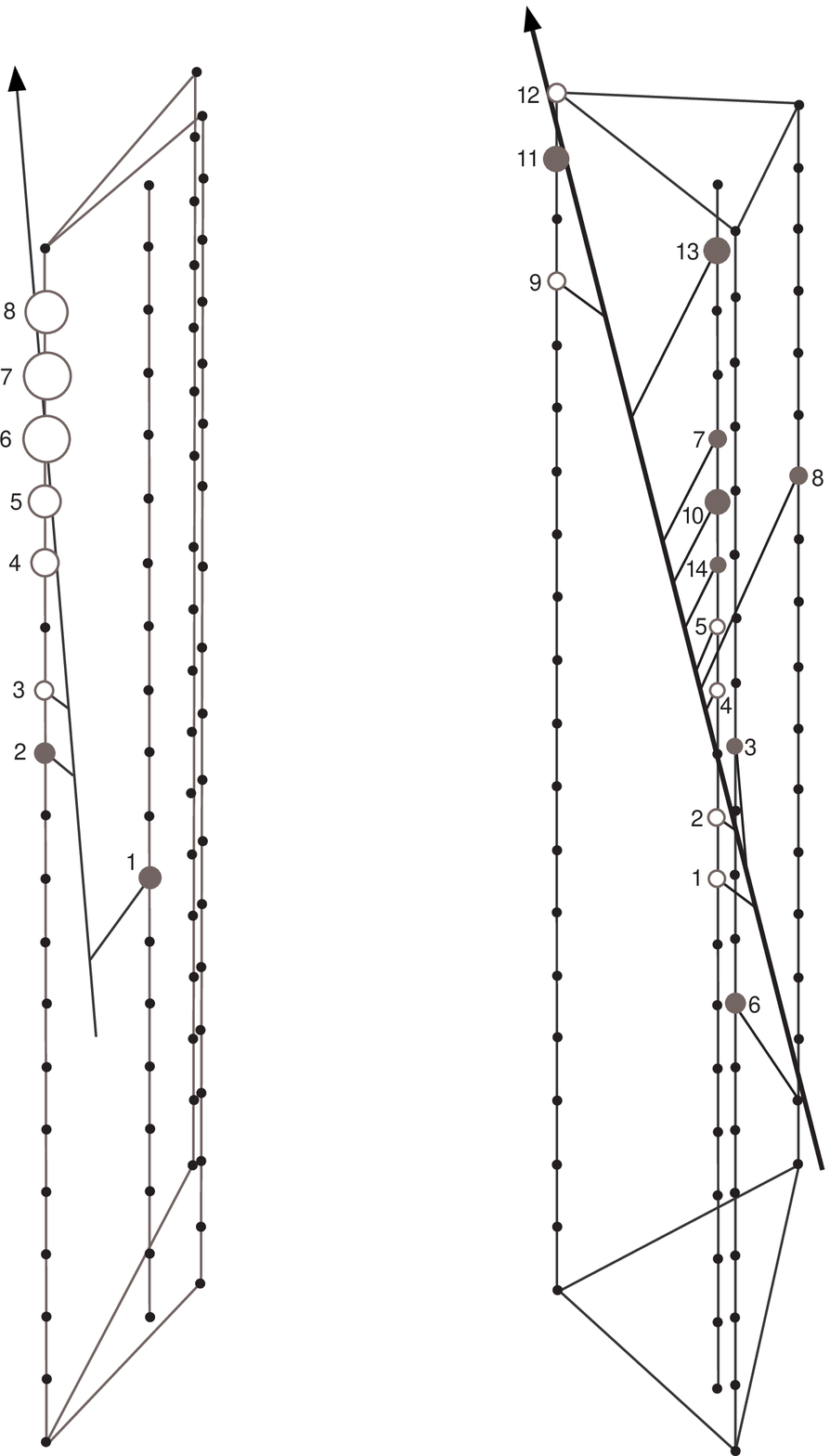,height=15.5cm}}
\caption[2]{\small
The two experimental events reconstructed as upward muons.
{\it left:} ID 8427997, {\it right:} ID 4706870. 
The line with an arrow symbolizes the fitted muon track, the lines
from this track to the OMs indicate light pathes. 
The amplitudes are proportional to the size of the OMs.
The numbering of the
OMs refers to the time order in which they are hit.
}
\label{twoevents}
\end{figure}

Fig.~\ref{ndirect} shows the distribution of the number of direct hits,
$N_{dir}$(15),
%(time residual in the interval [-15\,nsec,\,+15\,nsec]) 
of all events passing these cuts. 
The highest
cut in $N_{direct}$ survived by {\it any} experimental event
is $N_{direct} \ge 6$. The two surviving events are shown 
in fig.~\ref{twoevents}.
%Table~\ref{events} gives their parameters.
The Monte-Carlo
expectation for upward muons from atmospheric neutrinos
is 2.8 events, with an uncertainty of a factor 2, mostly due to 
uncertainties in the sensitivity of the detector after all
cuts.

\bigskip

{\large \it Analysis 2}
%\subsection{Standard Reconstruction}

%We used the same full experimental set as in the previous analysis,
The $3.5 \cdot 10^8$ experimental events were  
%This sample could be 
compared to $3.5 \cdot 10^6$ MC events from atmospheric down-going
muons which 
%represent potential fake events and 
correspond to 2 days 
effective live time. The MC data set for upward muons from 
atmospheric neutrino interactions \cite{Nutomu} consists of $2.5 \cdot 10^3$
events triggering AMANDA-B4 -- corresponding to 1.7 years
effective live time.

In order to separate neutrino induced upward muons, 
we applied a number of successively tightened cuts in the 
variables defined in section 6.4. 
This procedure 
reduced the experimental sample to the expected  
signal sample after the following cuts:

\begin{enumerate}

\item
reconstructed zenith angle $\theta > 120^o$,

\item 
speed of the line fit $ 0.15 <  |\vec{v}| < 1$ m/nsec,

\item
"time" likelihood  $\log({\cal L}_{time})/(N_{hit}-5) >  -10  $
(i.e. normalizing to the degrees of freedom instead of the
the number of hit PMTs),
 
\item 
"hit" likelihood  $\log({\cal L}_{hit})/(N_{hit}-5) > -8$,

\item 
number of direct hits for 25 nsec window, $N_{dir}(25) \ge 5 $, 

\item 
number of direct hits for 75 nsec  window, $N_{dir}(75) \ge 9 $, 

\item
projected length of direct hits for 25 nsec window, $L_{dir}(25) > 200$\,m,

\item
absolute value of the vertical coordinate of the center of gravity  
$ |z_{COG}| < 90$m \\
(with the center of the detector defining the origin of the coordinate
system).

\end{enumerate}

Three events of the experimental sample passed these cuts, 
corresponding to a suppression
factor of  $8.9 \cdot 10^{-9}$.  The passing rate for MC upward moving
muons from atmospheric neutrinos is 1.3 \%   which corresponds
to 4.0 events in 156 days. The corresponding enrichment
factor is $0.013/(8.9 \cdot 10^{-9}) \approx 1.5 \cdot 10^6$.
One of the three experimental events was identified also
in the search from the previous subsection. A second event 
with $N_{dir} = 5$ passes
all cuts of the previous analysis, with the exception of
the $N_{dir}$ cut.  

In order to check how well the parameter distributions of the
events agree with what one expects for atmospheric neutrino
interactions, and how well they are separated from the
rest of the experimental data, we relaxed two cuts at a time 
(retaining the rest) and  inspected the distribution in the
two "free" variables.

\begin{figure}[htbp]
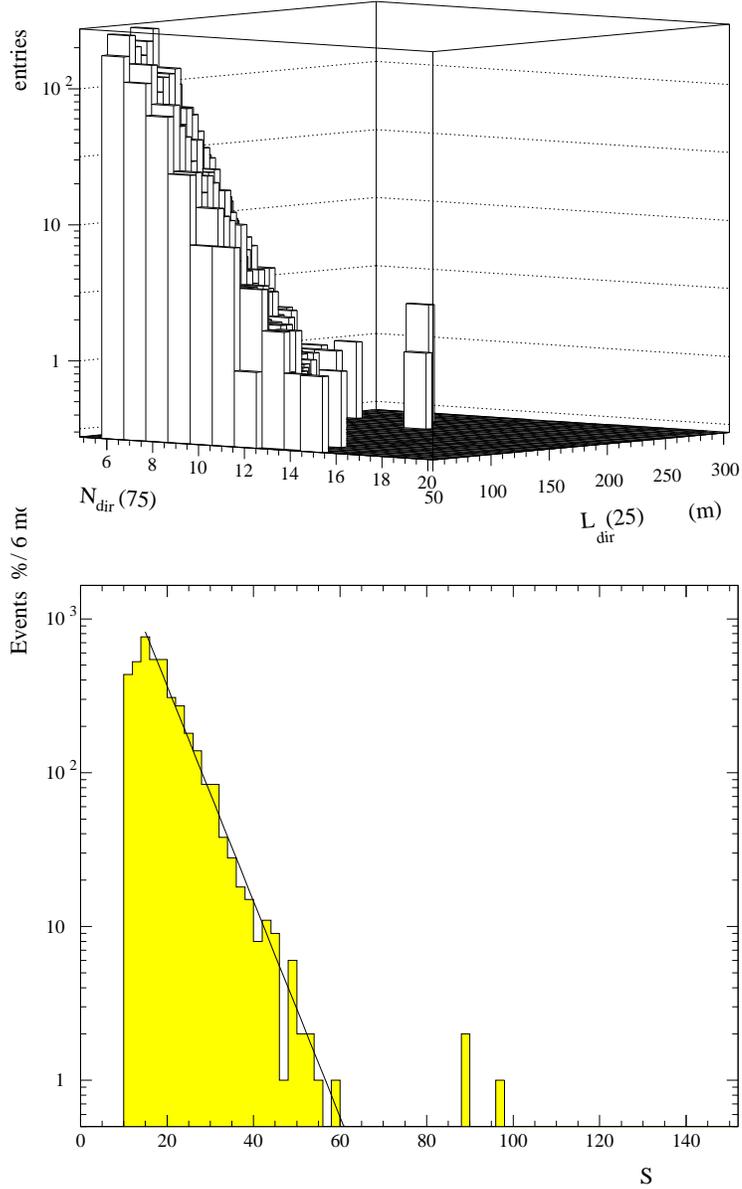

\centering
\mbox{\epsfig{file=96_sepa_3d.epsi,width=9.7cm}}
\mbox{\epsfig{file=96_sepa_1d.epsi,width=9.7cm}}
\caption[2]{\small
{\it Top } --  distribution in parameters $L_{dir}$(25)
vs. $N_{dir}$(75),  
{\it bottom:} distribution in the "combined" parameter
$S = N_{dir}$(75) $\cdot L_{dir}$(25) / 20.
The cuts applied to the event sample include
all cuts with the exception of cuts 6 and 7.
}
\label{sepa}
\end{figure}

Fig.~\ref{sepa} shows the distribution in $L_{dir}$(25)
and $N_{dir}$(75). The three events
passing {\it all} cuts are  separated from
the bulk of the data. At the bottom of fig.~\ref{sepa},
the data are plotted
versus a combined parameter, 
$  S = (N_{dir}$(75)-2) $\cdot L_{dir}$(25)/20.
In this parameter, the  data exhibit
a nearly exponential decrease. Assuming the decrease of the
background dominated events to continue at higher $ S $ values,
one can calculate the probability that the separated events are
fake events. The probability to observe one event
at $S \ge 70$ is 15\%, the probability to
observe 3 events is only $6 \cdot 10^{-4}$.

\begin{figure}[htbp]
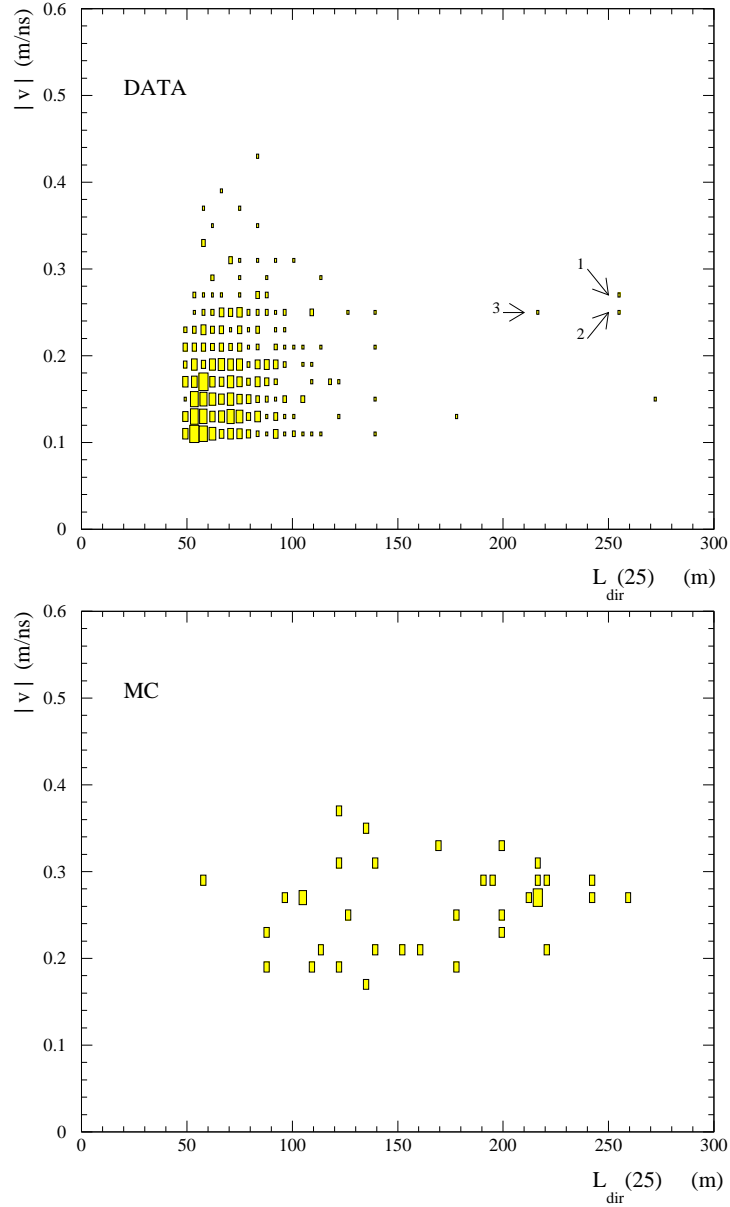

\centering
\mbox{\epsfig{file=96_velo_data.epsi,width=9.5cm}}
\mbox{\epsfig{file=96_velo_signal.epsi,width=9.5cm}}
\caption[2]{\small
Distribution in parameters $|\vec{v}|$ vs. $L_{dir}$(25),
after application of all cuts with the exception of 
cuts 2 and 7, which have been relaxed.
{\it top:} experimental data,  
{\it bottom:} signal Monte Carlo sample
}
\label{velo}
\end{figure}

Fig.~\ref{velo} shows the distribution when
$|\vec{v}|$  and $L_{dir}$(25) are relaxed. The 3 events are marked by
arrows. There is one additional event at high 
 $L_{dir}$(25), which, however, has a somewhat too small
$|\vec{v}|$. The 3 events fall into the
region populated by MC generated atmospheric neutrino events
passing the same cuts (bottom of fig.~\ref{velo}).
We attribute the lack of 
experimental events between  $L_{dir}$(25) $\sim$ 150--200
to statistical fluctuations.

Due to CPU limitation we could not check the
agreement between experimental data and atmospheric muon MC 
down to a  $8.9 \cdot 10^{-9}$ reduction. However, down to
a reduction of $10^{-5}$, the disagreement
does not exceed a factor of 3.
A less conservative estimate of the accuracy of the signal 
prediction can be obtained 
by replacing all dedicated  cuts for $\theta > 90^o$
by the complementary cuts for $\theta < 90^o$. We 
observed a better-than-10\% agreement between experimental data and 
MC after all cuts.  
In conclusion we estimate the uncertainty in the prediction of
upward muon neutrinos to be about a factor 2.

Table \#\ref{events} summarizes the characteristics of the 
neutrino candidates identified in the two analyses.

\begin{table}[ht]
\caption{
Characteristics  of the  events reconstructed as up-going muons
}
\begin{center}
\begin{tabular} {|c||c|c|c|c|} \hline
event ID $\rightarrow$ &  147\,742 & 4\,706\,879 & 2\,324\,428 & 8\,427\,905  \\ \hline \hline
%$\alpha$, m/nsec & 0.19 & 0.37  \\ \hline
%length, m & 295 & 182 \\ \hline
$N_{OM}$ & 13 & 14 & 15 &  8 \\ \hline
$N_{string}$ & 3 & 4 & 3 & 2 \\ \hline
log({\cal L}/$(N_{hit}-5))$ & -8.3 & -8.5  & -8.0  & -11.2 \\ \hline
$\theta_{rec}$, degrees & 168.7 & 165.9 & 166.7 & 175.4 \\ \hline
$\phi_{rec}$, degrees & 45.8 & 274.2 & 194.1 & -- \\ \hline 
\end{tabular}
\end{center}
\label{events}
\end{table}

We conclude that tracks reconstructed as up-going
are found at a rate consistent with that
expected for  atmospheric neutrinos. The three events
found in the second analysis
are well separated from background proving that, even
with a detector as small as AMANDA-B4, neutrino
candidates can be separated within a limited zenith
angle interval. 
Meanwhile, a few tens of clear neutrino events have
been identified with the more powerful AMANDA-B10 telescope.
They will be the subject of a forthcoming paper.

%\newpage

\section{Conclusions \label{conclusion}}

We have described the design, operation, calibration and selected
results of the prototype neutrino telescope AMANDA-B4 at the 
South Pole. 

The main results  can be summarized as follows:
\begin{itemize}
\item AMANDA-B4 consisting of 80 optical modules 
 (+ 6 OMs for technology tests) on 4 strings has
      been deployed at depths between
      1.5 and 2.0 km in 1996. Seven of the OMs failed during
      refreezing.
%      The six deepest PMTs have been used only for technology tests.
      We have developed  reliable drilling and
      instrumentation procedures allowing deployment of a 2 km deep
      string in less than a week. In the mean time the detector has
      been upgraded to 302 (AMANDA-B4, 1997) and 424 (1998) 
      optical modules.

\item The ice properties between 1.5 and 2.0 km are superior to those at
      shallow depths. The absorption length is about 95\,m and
      the effective scattering length about 24\,m.

\item The original calibration 
      accuracy reached for geometry and timing 
      of AMANDA-B4 was about 2 m and 7 nsec, respectively. With the
      upgrade to 10 strings, these values have been improved to 
      0.5~-~1.0 m and 5 nsec.

\item We have developed proper methods for track reconstruction
      in a medium with non-negligible scattering.
      With tailored quality cuts, the remaining badly reconstructed 
      tracks can be removed. The quality of the reconstruction
      and the efficiency of the cuts improve considerably with
      increasing size of the array. 

\item Geometry and tracking accuracy of AMANDA can be calibrated with
      surface air shower detectors. The mismatch  between showers
      detected in the SPASE air shower array and muons detected
      with AMANDA is about 4 degrees, in agreement with
      Monte Carlo estimates of the angular accuracy.

\item The measured angular spectrum of the intensity of 
      atmospheric muons is in good agreement with other 
      experiments and with model calculations.

\item First neutrino candidates have been separated with AMANDA-B4.
      The identification of upward muon candidates with an
      array of only 73 operating 8-inch PMTs is a demonstration that deep
      antarctic ice is an adequate medium for doing 
      neutrino astronomy.

\end{itemize}

Amanda-B4 is a first step towards a large neutrino telescope at the
South Pole. A ten-string array, AMANDA-B10, has been taking data
since 1997. Presently, B10 data are analyzed, and  tens of clear 
neutrino candidates have been extracted, with
a threshold of typically 50 GeV. 
The construction of AMANDA-II, a 30\,000 m$^2$ array,
is underway. The long-term goal of the
collaboration is a cube kilometer detector,
ICECUBE.

\section{Acknowledgments}

This research was supported by 
the U.S. National Science Foundation, Office of Polar Programs
and Physics Division,
the University of Wisconsin Alumni Research Foundation,
the U.S. Department of Energy,
the U.S. National Energy Research Scientific
Computing Center, 
the Swedish Natural Science Research Council,
the Swedish Polar Research Secretariat,
the Knut and Alice Wallenberg Foundation, Sweden,
and the Federal Ministery for Education and Research, Germany.
C.P.H. acknowledges the support of the
European Commission through TMR contract No. ERBFMBICT91551.

We thank the 
Polar Ice Coring Office, PICO, for bore hole
drilling, and the Antarctic Support Associates, ASA, 
as well as the staff of the
Amundsen Scott station for support and assistance.
We gratefully acknowledge help from the 
SPASE collaboration, Leeds University, and the
U.K. Particle Physics and Astrophysics Research
Council.

\end{document}